\documentstyle[11pt]{article}

\def\theequation{\arabic{section}.\arabic{equation}}

\renewcommand{\theequation}{\thesection.\arabic{equation}}

\global\arraycolsep=1pt
\input epsf
\input{tcilatex}
\begin{document}

\font\cmss=cmss10 \font\cmsss=cmss10 at 7pt \hfill \hfill IFUP-TH/02-44


\vspace{10pt}

\begin{center}
{\Large {\bf \vspace{10pt}ABSENCE\ OF\ HIGHER DERIVATIVES IN\ THE\
RENORMALIZATION\ OF PROPAGATORS\ IN\ QUANTUM\ FIELD\ THEORIES\ WITH\
INFINITELY\ MANY\ COUPLINGS}}

\bigskip \bigskip

{\sl Damiano Anselmi}

{\it Dipartimento di Fisica E. Fermi, Universit\`{a} di Pisa, and INFN}
\end{center}

\vskip 2truecm

\begin{center}
{\bf Abstract}
\end{center}

{\small I study some aspects of the renormalization of quantum field
theories with infinitely many couplings in arbitrary space-time dimensions.
I prove that when the space-time manifold admits a metric of constant
curvature the propagator is not affected by terms with higher derivatives.
More generally, certain lagrangian terms are not turned on by
renormalization, if they are absent at the tree level. This restricts the
form of the action of a non-renormalizable theory, and has applications to
quantum gravity. The new action contains infinitely many couplings, but not
all of the ones that might have been expected. In quantum gravity, the
metric of constant curvature is an extremal, but not a minimum, of the
complete action. Nonetheless, it appears to be the right perturbative
vacuum, at least when the curvature is negative, suggesting that the quantum
vacuum has a negative asymptotically constant curvature. The results of this
paper give also a set of rules for a more economical use of effective
quantum field theories and suggest that it might be possible to give
mathematical sense to theories with infinitely many couplings at high
energies, to search for physical predictions. }

\vspace{4pt}

\vskip 1truecm

\vfill\eject

\section{Introduction}

The quantization of gravity is still elusive. The removal of divergences of
quantum gravity is possible only in the presence of infinitely many
independent coupling constants and it is hard, although not impossible in
principle, to find computable quantities and make physical predictions about
the high-energy behavior of the theory. On the other hand, the strenuous
efforts spent in the recent years to search for approaches ``beyond quantum
field theory'' have not produced significant breakthroughs. In the absence
of convincing alternatives, it is wiser to investigate the problem of
quantum gravity in the best established framework and get used to work with
infinitely many couplings.

The action 
\begin{equation}
-\frac{1}{\kappa ^{2}}\int \sqrt{g}\ \left( R-\Lambda \right)  \label{EH}
\end{equation}
is not positive definite, in the Euclidean framework. The complete action of
quantum gravity, however, is not (\ref{EH}), but contains infinitely many
terms besides (\ref{EH}), with arbitrarily high powers and derivatives of
the curvature tensors. In these circumstances, we need not worry about the
positive indefiniteness of (\ref{EH}). We can define the functional integral
formally, and perturbatively, relaxing the requirement of rigorous
convergence of the term-by-term functional integration.

Usually, non-renormalizable theories are used as effective low-energy
theories, because the number of parameters that are necessary to remove the
divergences is finite at low energies (but grows with the energy).

As opposed to an ``effective'' theory, a ``fundamental theory'' should be
well-defined and predictive at arbitrarily high energies. In the usual
sense, a fundamental theory is predictive only if it contains a finite
number of independent coupling constants. If it contains infinitely many
coupling constants, certain subclasses of correlation functions might still
depend only on a finite number of parameters, or a finite number of
functions of them. Then the problem is to identify such classes of
correlation functions.

In this paper I consider the theories with infinitely many coupling
constants and show that some questions are well posed, and can be answered.
The purpose of this research is to construct a class of theories that lie at
an intermediate stage between effective and fundamental quantum field
theory, e.g. quantum gravity with infinitely many parameters. The final goal
is to use these theories to derive {\it some} physical predictions beyond
the low-energy domain and possibly classify {\it which} physical predictions
can be derived from these theories.

The first step in the task of giving mathematical sense to quantum field
theories with infinitely many parameters at arbitrary energies is to show
that a unitary propagator is not driven by renormalization into a
non-unitary (typically, higher-derivative) propagator. The reason to worry
that this might happen is that the non-renormalizability of the theory can
potentially generate all sorts of counterterms, including those that can
affect the propagator with undesirable higher derivatives. In this paper I
show that this problem can be solved when the space-time manifold admits a
metric of constant curvature. Furthermore, it is possible to screen the
terms of the lagrangian and prove that, for example, a whole class of terms
is not turned on by renormalization, if it is absent at the tree level.

\bigskip

In the absence of matter, quantum gravity is one-loop finite \cite
{thooftveltman}: the one-loop divergent terms of the form $R^{2}$, $R_{\mu
\nu }^{2}$ can be removed with a redefinition of the metric tensor. However,
when gravity is coupled to matter, the counterterms $R^{2}$, $R_{\mu \nu
}^{2}$ are non-trivial and can in principle be responsible for the
appearance of unphysical singularities in the graviton propagator. The
unphysical singularities of the propagator are relevant only at high
energies and can be ignored if the perturbative expansion is truncated to a
finite power of the energy \cite{wein}, e.g. in the framework of effective
field theory. However, this is a way to ignore the problem, rather than
solve it. The final goal of perturbation theory is the resummation of the
series expansion, at least in suitable correlation functions and physical
quantities. Fortunately, in the presence of matter, the undesired
counterterms $R^{2}$, $R_{\mu \nu }^{2}$ can be traded for renormalizations
of the vertex couplings and a redefinition of the metric tensor, namely
there exists a subtraction scheme where the graviton propagator is not
affected by higher derivatives. Moreover, this fact generalizes to every
order in the perturbative expansion and in arbitrary space-time dimensions,
under some mild assumptions. The absence of unphysical singularities in the
propagators is a necessary condition for unitarity, although not sufficient.
It is worth saying that this result does not ensure that, for example, the
two-point functions of the fields are free of Landau poles (Landau poles are
difficult to treat also in the context of renormalizable theories).

\bigskip

To fix some basic terminology, I distinguish two subclasses of counterterms:
the quadratic counterterms and the vertex counterterms. By quadratic
counterterms I mean the counterterms that, expanded perturbatively, have
contributions quadratic in the quantum fluctuations of the fields. By vertex
counterterms I mean the counterterms that, expanded perturbatively, have no
contributions quadratic and linear in the quantum fluctuations of the
fields. Let us consider four-dimensional pure quantum gravity without a
cosmological constant in detail, 
\begin{equation}
{\cal L}=-\frac{1}{\kappa ^{2}}\sqrt{g}R(x).  \label{qg}
\end{equation}
The one-loop divergences \cite{thooftveltman} 
\begin{equation}
\frac{1}{8\pi ^{2}\varepsilon }\sqrt{g}\left( \frac{1}{120}R^{2}+\frac{7}{20}%
R_{\mu \nu }R^{\mu \nu }\right)  \label{cou}
\end{equation}
can be removed in two different ways.

$i$) New coupling constants are added and the lagrangian (\ref{qg}) is
modified into 
\begin{equation}
{\cal L}^{\prime }=\frac{1}{\kappa ^{2}}\sqrt{g}\left( -R(x)+\alpha _{{\rm B}%
}R^{2}+\beta _{{\rm B}}R_{\mu \nu }R^{\mu \nu }\right) ,  \label{hdqg}
\end{equation}
The divergences (\ref{cou}) are removed defining the bare parameters 
\[
\alpha _{{\rm B}}=\left( \alpha -\frac{1}{8\pi ^{2}\varepsilon }\frac{1}{120}%
\right) \mu ^{-\varepsilon },\qquad \qquad \beta _{{\rm B}}=\left( \beta -%
\frac{1}{8\pi ^{2}\varepsilon }\frac{7}{20}\right) \mu ^{-\varepsilon }. 
\]

The replacement (\ref{qg}) $\rightarrow $ (\ref{hdqg}) changes the theory
into higher-derivative quantum gravity, which is renormalizable, but not
unitary \cite{stelle}. The renormalizability is due to the resummation of
the power series of $\alpha $ and $\beta $ in the graviton propagator. The
unphysical singularities are of orders $1/\alpha $ and $1/\beta $.

Once the couplings $\alpha $ and $\beta $ are introduced, there are energy
domains where the values of $\alpha $ and $\beta $ cannot be considered
small, and the modified theory (\ref{hdqg}) of quantum gravity, as a
fundamental theory, violates unitarity at high energies.

$ii$) We observe that (\ref{cou}) vanish using the field equations of (\ref
{qg}) and remove (\ref{cou}) with field redefinitions of the form 
\begin{equation}
g_{\mu \nu }\rightarrow g_{\mu \nu }+\frac{\kappa ^{2}}{8\pi ^{2}\varepsilon 
}\frac{1}{20}\left( -7R_{\mu \nu }+\frac{11}{3}g_{\mu \nu }R\right) ,
\label{redef}
\end{equation}
up to two- and higher-loop contributions. The theory remains
non-renormalizable, but unitary. In this case, the couplings $\alpha $ and $%
\beta $ are called ``inessential'' \cite{wein}.

The explicit computation of the one-loop divergences is actually unnecessary
to prove finiteness, because the one-loop divergences are a linear
combination of terms quadratic in the curvature tensors, namely 
\begin{equation}
R_{\mu \nu \rho \sigma }R^{\mu \nu \rho \sigma },\qquad R_{\mu \nu }R^{\mu
\nu },\qquad R^{2},  \label{form}
\end{equation}
but the identity 
\begin{equation}
\sqrt{g}\left( R_{\mu \nu \rho \sigma }R^{\mu \nu \rho \sigma }-4R_{\mu \nu
}R^{\mu \nu }+R^{2}\right) =\text{total derivative},  \label{identi}
\end{equation}
can be used to convert $R_{\mu \nu \rho \sigma }R^{\mu \nu \rho \sigma }$
into the sum $4R_{\mu \nu }R^{\mu \nu }-R^{2}$, which is proportional to the
vacuum field equations and can be eliminated with a redefinition of the
metric tensor. However, (\ref{identi}) is true only in four dimensions,
because to prove (\ref{identi}) it is necessary to use that a completely
antisymmetric tensor with more than four indices vanishes. It is therefore
natural to wonder whether in higher dimensions gravity is always driven to
higher-derivative gravity by renormalization. I am going to show that this
is not the case, because suitable generalizations of (\ref{identi}) do exist.

Finiteness of one-loop pure quantum gravity is a coincidence, spoiled by the
presence of matter \cite{thooftveltman}. Moreover, Goroff and Sagnotti \cite
{sagnotti} proved that pure quantum gravity is not finite at the second loop
order, but there appears a counterterm proportional to 
\begin{equation}
R_{\mu \nu }^{\rho \sigma }R_{\alpha \beta }^{\mu \nu }R_{\rho \sigma
}^{\alpha \beta },  \label{gs}
\end{equation}
which cannot be reabsorbed by means of field redefinitions. Finally, in
higher dimensions pure quantum gravity is not even finite at the one-loop
order.

\bigskip

As anticipated above, the identity (\ref{identi}) admits a number of
generalizations. If the metric is expanded around a flat background, 
\[
g_{\mu \nu }=\delta _{\mu \nu }+\phi _{\mu \nu }, 
\]
the dimension-independent formula 
\[
R_{\mu \nu \rho \sigma }=\frac{1}{2}\left( \partial _{\rho }\partial _{\nu
}\phi _{\mu \sigma }-\partial _{\rho }\partial _{\mu }\phi _{\nu \sigma
}-\partial _{\sigma }\partial _{\nu }\phi _{\mu \rho }+\partial _{\sigma
}\partial _{\mu }\phi _{\nu \rho }\right) +{\cal O}(\phi ^{2}) 
\]
implies that in the combination 
\[
\sqrt{g}\ {\rm G}\equiv \sqrt{g}\left( R_{\mu \nu \rho \sigma }R^{\mu \nu
\rho \sigma }-4R_{\mu \nu }R^{\mu \nu }+R^{2}\right) 
\]
the sum of the quadratic terms in $\phi _{\mu \nu }$ is a total derivative
in every space-time dimension. This ensures that the integral $\int \sqrt{g}%
\ {\rm G}$ is a vertex, 
\begin{equation}
\int \sqrt{g}\ {\rm G}={\cal O}(\phi ^{3}),  \label{atom}
\end{equation}
and therefore the divergences of the form (\ref{form}) do not affect the
propagator with higher derivatives. It is easy to prove that the right-hand
side of (\ref{atom}) does not vanish in dimensions greater than four.

This observation can be generalized and applied to show that renormalization
protects unitarity even in the presence of infinitely many couplings. I
prove that it is possible to remove the higher-derivative quadratic
counterterms by means of field redefinitions in two cases: in the absence of
a cosmological term and when the space-time manifold admits a metric of
constant curvature. This property holds also in the presence of matter and
in arbitrary dimension greater than two, for theories containing fields of
spin 0, 1/2, 1, 3/2 and 2.

I believe that the true meaning of the identity (\ref{identi}) and its
generalizations, such as (\ref{atom}) and (\ref{bibi}), is the absence of
higher-derivative corrections to the propagators in theories with infinitely
many couplings, rather than the finiteness of special truncations of some
theories.

When the space-time manifold does not admit a metric of constant curvature,
a more general version of the theorem ensures that a certain class of terms
is not turned on by renormalization, if it is absent at the tree level. This
allows us to write an explicit form of the lagrangian of quantum gravity,
which does contain infinitely many parameters, but not all of the ones that
we might have expected.

If the metric is expanded around a vacuum metric with non-constant
curvature, the graviton propagator does contain higher derivatives. This
happens also in non-renormalizable theories of fields with spin 0, 1/2 and
1, expanded around non-trivial backgrounds, for example istantons. The issue
of unitarity around nontrivial backgrounds is delicate and, in the case of
instantons, the matter is further complicated by the integration over the
instanton moduli space. Here I do not prove sufficient conditions for
unitarity, but a simple theorem ensuring that under the mentioned
assumptions renormalization does not generate unphysical poles in the
perturbative propagator.

I also show that the results of this paper priviledge a spacetime manifold
admitting a metric of constant curvature and therefore suggest that the
quantum vacuum has an asymptotically constant curvature. The metric of
constant curvature is an extremal, although not a minimum of the complete
action. Nevertheless, it seems to be the correct perturbative vacuum for the
calculations in quantum gravity, at least when the constant curvature is
negative.

\bigskip

The paper is organized as follows. In section 2 I briefly discuss the
relation between higher derivatives, unphysical singularities of the
propagator and normalization conditions of the couplings in quantum field
theory. In section 4 I treat the non-renormalizable theories of fields with
spin 0, 1/2 and 1, while in section 5 I treat the spin-3/2 fields. In
sections 6 and 7 I study quantum gravity without and with a cosmological
constant, respectively, and write the complete action of quantum gravity. In
section 8 I make some observations about the perturbative vacuum and the
vacuum of quantum gravity. Section 9 collects some conclusions.

\section{Higher derivatives and unitarity}

When the quadratic part of a lagrangian contains higher derivatives the
initial conditions in classical mechanics and the normalization conditions
for the two-point function in quantum field theory are non-standard and
reveal the presence of ghosts.

\bigskip

{\bf Classical mechanics.} In classical mechanics the presence of higher
(time) derivatives implies that the solution of the equations of motion is
not uniquely determined by the initial positions and velocities of the
particles. For example, the lagrangian

\begin{equation}
{\cal L}=\frac{m}{2}\left( \frac{{\rm d}q}{{\rm d}t}\right) ^{2}+\frac{%
\alpha }{2}\left( \frac{{\rm d}^{2}q}{{\rm d}t^{2}}\right) ^{2}-V(q)
\label{kine}
\end{equation}
generates an equation of motion containing the fourth time derivative of $q$%
, 
\begin{equation}
m\frac{{\rm d}^{2}q}{{\rm d}t^{2}}-\alpha \frac{{\rm d}^{4}q}{{\rm d}t^{4}}+%
\frac{\partial V(q)}{\partial q}=0,  \label{ecu}
\end{equation}
whose solution is unique if the position, velocity, acceleration and the
derivative of the acceleration at a reference time $t_{0}$ are specified: 
\[
q(t_{0})=q_{0},\qquad \frac{{\rm d}q}{{\rm d}t}(t_{0})=v_{0},\qquad \frac{%
{\rm d}^{2}q}{{\rm d}t^{2}}(t_{0})=a_{0},\qquad \frac{{\rm d}^{3}q}{{\rm d}%
t^{3}}(t_{0})=\dot{a}_{0}. 
\]
This means that the theory does not propagate just one field, but more
fields. However, the additional fields are ghosts. This can be quickly
viewed as follows.

The potential and kinetic terms of (\ref{kine}) are positive definite, if $%
\alpha >0$ and $V$ is positive. If we introduce an additional variable $Q$
and rewrite the lagrangian without higher derivatives, for example 
\[
{\cal L}^{\prime }=\frac{m}{2}\left( \frac{{\rm d}\widetilde{q}}{{\rm d}t}%
\right) ^{2}-\frac{\alpha ^{2}}{2m}\left( \frac{{\rm d}Q}{{\rm d}t}\right)
^{2}-\frac{\alpha }{2}Q^{2}-V\left( \widetilde{q}+\frac{\alpha }{m}Q\right)
, 
\]
we obtain a theory whose equations of motion are equivalent to those of (\ref
{ecu}), with $\widetilde{q}=q-\alpha Q/m$. The new potential is still
positive definite, but the new kinetic term is not positive definite.

\bigskip

{\bf Quantum field theory.} Consider for example the $\varphi ^{4}$ theory 
\begin{equation}
{\cal L}=\frac{1}{2}\left( \partial _{\mu }\varphi \right) ^{2}+\frac{m^{2}}{%
2}\varphi ^{2}+\frac{\lambda }{4!}\varphi ^{4}.  \label{theo}
\end{equation}
The physical constants and the normalization of the field $\varphi $ have to
be fixed at a reference energy scale $\mu $ by means of suitable {\it %
normalization conditions}, such as 
\begin{eqnarray}
\left. \frac{\partial ^{2}\Gamma ^{(2)}[p]}{\partial p^{2}}\right|
_{p^{2}=\mu ^{2}} &=&1,\qquad  \label{norma1} \\
\left. \Gamma ^{(2)}[p]\right| _{p^{2}=\mu ^{2}}=\mu ^{2}+m^{2}, &&\qquad
\left. \Gamma ^{(4)}[p_{1},p_{2},p_{3}]\right| _{S}=\lambda .  \label{norma2}
\end{eqnarray}
where $S$ denotes the symmetric condition $p_{1}^{2}=p_{2}^{2}=p_{3}^{2}=\mu
^{2}$, $s=t=u=4\mu ^{2}/3$, and $s=\left( p_{1}+p_{2}\right) ^{2}$, $%
t=\left( p_{1}+p_{3}\right) ^{2}$, $u=\left( p_{2}+p_{3}\right) ^{2}$, as
usual. Here $m$ and $\lambda $ denote the renormalized parameters, but might
not be the physical mass and the physical coupling constant. The first
condition (\ref{norma1}) is conventional and fixes the normalization of the
field $\varphi $, which has no physical significance. The second and third
conditions (\ref{norma2}) fix $m$ and $\lambda $. It is imagined that the
quantities appearing in the left-hand sides of (\ref{norma2}) are determined
by some experimental observation at the energy scale $\mu $. Since $\left.
\Gamma ^{(2)}[p]\right| _{p^{2}=\mu ^{2}}$ and $\left. \Gamma
^{(4)}[p_{1},p_{2},p_{3}]\right| _{S}$ depend only on the unknowns $m$ and $%
\lambda $, the normalization conditions (\ref{norma2}) fix $m$ and $\lambda $
and therefore the theory.

The theories containing infinitely many couplings need infinitely many
normalization conditions. Let us focus on the normalization conditions
associated with the two-point functions, which generalize (\ref{norma1}) and
the first of (\ref{norma2}). When the propagator contains higher
derivatives, 
\[
{\cal L}=\frac{1}{2}\left( \partial _{\mu }\varphi \right) ^{2}+\frac{m^{2}}{%
2}\varphi ^{2}+\frac{1}{2}\sum_{n=1}^{\infty }\alpha _{n}(\partial _{\mu
}\varphi )\Box ^{n}(\partial _{\mu }\varphi )+\text{vertices,} 
\]
the normalization conditions for $\Gamma ^{(2)}[p]$ read 
\begin{eqnarray}
\left. \left( \frac{\partial ^{2}}{\partial p^{2}}\right) ^{k}\Gamma
^{(2)}[p]\right| _{p^{2}=\mu ^{2}} &=&\sum_{n=k-1}^{\infty }\alpha
_{n}(-1)^{n}\frac{(n+1)!}{(n-k)!}\mu ^{2(n+1-k)},\qquad  \label{norma} \\
\left. \Gamma ^{(2)}[p]\right| _{p^{2}=\mu ^{2}} &=&\mu
^{2}+m^{2}+\sum_{n=1}^{\infty }\alpha _{n}(-1)^{n}\mu \Sp 2n+2  \\  \endSp .
\nonumber
\end{eqnarray}
for $k=1,2,\ldots $ and $\alpha _{0}=1$. The need for normalization
conditions with $k>1$ reveals the presence of ghosts.

A necessary condition for unitarity is that no normalization conditions (\ref
{norma}) with $k>1$ are necessary to determine the theory. The theorem
proven in this paper ensures this, namely that the couplings $\alpha _{n}$
are inessential and the most general form of a non-renormalizable lagrangian
is 
\[
{\cal L}=\frac{1}{2}\left( \partial _{\mu }\varphi \right) ^{2}+\frac{m^{2}}{%
2}\varphi ^{2}+\text{vertices.} 
\]
The normalization conditions for $\Gamma ^{(2)}[p]$ are just 
\[
\left. \frac{\partial ^{2}\Gamma ^{(2)}[p]}{\partial p^{2}}\right|
_{p^{2}=\mu ^{2}}=1,\qquad \left. \Gamma ^{(2)}[p]\right| _{p^{2}=\mu
^{2}}=\mu ^{2}+m^{2}. 
\]
The result is trivial for scalar fields and fermions (see below), which I
consider just for illustrative purposes. The conclusions apply to the most
general theory of spin-0, -1/2, -1, -3/2 and -2 fields in arbitrary
dimension greater than two, in the absence of a cosmological constant and
when the space-time manifold admits a metric of constant curvature. A
generalization that does not need these restrictions exists, and will be
discussed later.

\section{Removal of divergences and higher derivatives}

\setcounter{equation}{0}

The purpose of this section and the next one is to show that

{\it in a non-renormalizable quantum field theory of fields of spin 0, 1/2,
1, 3/2 and 2 with no cosmological term, higher-derivative quadratic terms
are not turned on by renormalization, if they are absent at the tree level.}

On the contrary, if a theory has a higher-derivative propagator at the
classical level, infinitely many new higher-derivative quadratic
counterterms, not removable using the field equations, can be generated by
renormalization.

I begin recalling some basic facts about the removal of divergences in gauge
theories and then derive a sort of uniqueness property for the propagator.

\bigskip

{\bf Algorithm for the removal of divergences}. As usual, I proceed
inductively in the perturbative expansion. At the $n$th step, the
``classical'' action $S_{n}[\Phi ,K,\lambda ]$ is assumed to contain
appropriate counterterms so that the quantum action $\Gamma _{n}[\Phi
,K,\lambda ]$ is convergent to the order $\hbar ^{n}$ included. Here $\Phi $
denote collectively the fields, $K$ are the BRS sources and the sources
coupled to the composite operators, and $\lambda $ are the coupling
constants. The theorem of locality of the counterterms ensures that the
order-$\hbar ^{n+1}$ divergent part $\Gamma _{n+1\ {\rm div}}$ of $\Gamma
_{n}$ is local, since, by inductive assumption, the subdivergences of the
graphs contributing to $\Gamma _{n+1\ {\rm div}}$ have been removed.

The algorithm for the removal of divergences is made of two ingredients (see
for example \cite{me,me2}): field and source redefinitions (``canonical''
transformations) and redefinitions of the coupling constants. At the $n$th
step, the divergent terms $\Gamma _{n+1\ {\rm div}}$ can be removed \cite
{me,me2} with suitable order-$\hbar ^{n+1}$ divergent redefinitions of the
fields, sources and coupling constants, 
\[
\Phi \rightarrow \Phi +\delta _{n+1}\Phi ,\qquad K\rightarrow K+\delta
_{n+1}K,\qquad \lambda \rightarrow \lambda +\delta _{n+1}\lambda , 
\]
such that the classical action $S_{n+1}[\Phi ,K,\lambda ]$%
\[
S_{n+1}[\Phi ,K,\lambda ]=S_{n}[\Phi +\delta _{n+1}\Phi ,K+\delta
_{n+1}K,\lambda +\delta _{n+1}\lambda ]=S_{n}[\Phi ,K,\lambda ]-\Gamma
_{n+1\ {\rm div}}+{\cal O}(\hbar ^{n+2}). 
\]
generates a quantum action $\Gamma _{n+1}[\Phi ,K,\lambda ]=\Gamma _{n}[\Phi
,K,\lambda ]-\Gamma _{n+1\ {\rm div}}+{\cal O}(\hbar ^{n+2})$ that is
convergent to the order $\hbar ^{n+1}$ included. Observe that, since $\delta
_{n+1}$ is of order $\hbar ^{n+1}$, at each inductive step only the first
order of the Taylor expansion of $S_{n}[\Phi +\delta _{n+1}\Phi ,K+\delta
_{n+1}K,\lambda +\delta _{n+1}\lambda ]$ in $\delta _{n+1}$ is relevant. The
higher orders of the Taylor expansion in $\delta _{n+1}$ contribute to $%
\Gamma _{m\ {\rm div}}$ at the subsequent steps of the iterative procedure,
i.e. for $m>n+1$. Therefore, in the arguments of the next sections we will
be concerned only with the first order of the Taylor expansion in the field
redefinitions.

The terms that cannot be reabsorbed by means of field redefinitions have to
be reabsorbed by means of redefinitions of the coupling constants. When the
classical action $S_{0}$ does not contain the necessary coupling constants $%
\lambda $, new coupling constants have to be introduced. If the theory is
not renormalizable, the divergences can be removed only at the price of
introducing infinitely many coupling constants.

The field redefinitions can be of two types: ``gauge-covariant'' field
redefinitions, which do not change the BRS\ transformations of the fields,
and field redefinitions that change the BRS transformations (usually, in a
very complicated way). The second type of field redefinitions greatly
complicate our discussion and although they can be dealt with using the
general renormalization algorithm of \cite{me,me2}, they can be avoided
using the background field method \cite{back}, which is indeed quite popular
in quantum gravity (see for example \cite{sagnotti}). The background field
method is equivalent to the choice of a specific class of gauge fixings.
General gauge invariance can be proved using the approach of \cite{me,me2},
and the physical results are of course the same.

\bigskip

We are mostly concerned with counterterms proportional to the field
equations, because they can be inductively removed by means of field
redefinitions. It is sufficient to isolate the first (quadratic)
contributions in the quantum fluctuations to the field equations, that is to
say the terms proportional to $\left( -\Box +m^{2}\right) \varphi $ for
scalars, the terms proportional to $\left( D\!\!\!\!\slash+m\right) \psi $
for fermions, the terms proportional to $D_{\mu }F_{\mu \nu }$ for vectors
and the terms proportional to the Ricci tensor and the Ricci curvature for
gravity. We have to show that every higher-derivative divergent term that is
quadratic in the quantum fluctuations multiplies an inessential coupling,
i.e. it can be removed with a vertex counterterm, plus a field redefinition.

Formally, the divergences $\Gamma _{{\rm div}}[\varphi ,\lambda ]$ of the
functional integral

\[
\int {\rm d}[\varphi ]\text{ }{\rm \exp }\left( -S[\varphi ,\lambda ]+\int
J\varphi ,\right) 
\]
where $\lambda $ denote the parameters of the theory (couplings and masses)
are inductively subtracted redefining the fields and the parameters, 
\[
\int {\rm d}[\varphi ]\text{ }{\rm \exp }\left( -S[\varphi ,\lambda ]+\Gamma
_{{\rm div}}[\varphi ,\lambda ]+\int J\varphi \right) =\int {\rm d}[\varphi ]%
\text{ }{\rm \exp }\left( -S[\widetilde{\varphi }(\varphi ),\widetilde{%
\lambda }]+\int J\varphi \right) =\text{finite}, 
\]
and $\widetilde{\varphi }$ is a complicated function of $\varphi $,
generically containing infinitely many terms. $\widetilde{\varphi }$ plays
the role of the bare field, while $\varphi $ is the renormalized field. The
result is that $S[\widetilde{\varphi }(\varphi ),\widetilde{\lambda }]$
keeps the same form as $S[\varphi ,\lambda ]$, that is to say no independent
higher-derivative quadratic term is turned on, if it is absent in $S[\varphi
,\lambda ]$.

\bigskip

{\bf Truncations.} At the practical level, the renormalization procedure
needs to be equipped with the definition of appropriate truncations of the
theory. Indeed, since the number of parameters is infinite, infinitely many
Feynman diagrams contribute at each order of the $\hbar $ expansion. In
general, infinitely many diagrams cannot be computed in one shot. To avoid
this difficulty, it is convenient to define truncated theories $\Gamma
^{(N)} $. The quantization of a non-renormalizable theory proceeds from the
low to the high energies, and is an expansion in $E\kappa $, where $E$ is
the energy scale of a physical process and $\kappa $ is a constant of
dimension $-1$ in units of mass. The truncated theory $\Gamma ^{(N)}$ is
defined as the theory where only the powers $\left( E\kappa \right) ^{m}$
with $m\leq N$ are kept. In the truncated theory finitely many lagrangian
terms and coupling constants need to be considered. Moreover, at each order
of the $\hbar $ expansion finitely many Feynman diagrams contribute and the
divergent terms of order $\left( E\kappa \right) ^{m}$ with $m>N$ need not
be removed.

In summary, we have two expansions contemporarily: the expansion in powers
of $\hbar $ and the expansion in powers of the energy. The fundamental
theory is the limit $N\rightarrow \infty $ of $\Gamma ^{(N)}$. A property of
the fundamental theory is a property of $\Gamma ^{(N)}$ for every $N$, not
depending on $N$. For example, the property that we are going to prove (the
absence of higher derivatives in the propagators) is a property of the
fundamental theory, since it is a property of $\Gamma ^{(N)}$ for every $N$.
On the other hand, the dependence on finitely many parameters is not a
property of the fundamental theory, because the number of parameters of $%
\Gamma ^{(N)}$ is finite, but depends on $N$ and becomes arbitrarily large
when $N$ becomes large.

For concreteness, in pure four-dimensional quantum gravity without a
cosmological constant $\kappa $ is the inverse Planck mass and the action of
the $2N$th truncated theory reads symbolically 
\[
S_{2N}=\frac{1}{\kappa ^{2}}\int \sqrt{g}\left( -R+\kappa ^{4}\lambda
_{3}R^{3}+\cdots +\kappa ^{2N-2}\lambda _{N}R^{N}\right) . 
\]
Higher powers of $R$ can be neglected in the $2N$th truncation. Phenomena at
energies above the Planck mass can be described (at least in principle) only
after appropriate resummations.

Finally, in the presence of a cosmological constant $\Lambda $, there is no $%
\Lambda $-independent definition of the ``typical energy'' $E$ of a physical
process and the $2N$th truncated theory is defined as the theory where only
the powers $\left( E\kappa \right) ^{p}\left( \Lambda \kappa ^{2}\right)
^{q} $ with $p+2q\leq 2N$ are kept.

\bigskip

{\bf ``Uniqueness'' of the propagator.} The theorem we want to prove can be
illustrated with a simple argument, which is almost a proof in itself. Let
us symbolically write the action $S$ and the field equations ${\rm E}_{\Phi
}=0$ as 
\[
S=-\frac{1}{2}\int \Phi Q_{0}\Phi +{\cal O}(\Phi ^{3}),\text{\qquad }{\rm E}%
_{\Phi }=\frac{\delta S}{\delta \Phi }=-Q_{0}\Phi +{\cal O}(\Phi ^{2}). 
\]
Here $\Phi $ denote the fields and $Q_{0}$ is the inverse propagator, which
is two-derivative for bosons, and one-derivative for fermions. We can focus
on gauge fields, since the statement is trivial in the case of scalars and
spin-1/2 fermions (see below). We assume that the gauge fields are expanded
around trivial vacuum configurations ($A_{\mu }=0$ for vectors, $\psi _{\mu
}=0$ for spin-3/2 fields and $g_{\mu \nu }=\delta _{\mu \nu }$ for the
metric tensor). Gauge invariance to the lowest order in the expansion around
the vacua reads $\delta _{0}A_{\mu }=\partial _{\mu }\Lambda $, $\delta
_{0}\psi _{\mu }^{\alpha }=\partial _{\mu }\epsilon ^{\alpha }$ and $\delta
_{0}\phi _{\mu \nu }=\partial _{\mu }\xi _{\nu }+\partial _{\nu }\xi _{\mu }$%
, where $\alpha $ is a Lorentz index and $\phi _{\mu \nu }=g_{\mu \nu
}-\delta _{\mu \nu }$. We have $\delta _{0}Q_{0}\Phi =0$.

The higher-derivative quadratic counterterms, which we write as 
\[
\Delta {\cal L}_{i}=-\frac{1}{2}\Phi Q_{i}\Phi , 
\]
where $Q_{i}$ is some polynomial in the derivatives, have to be invariant
under the lowest-order gauge transformations. Using the gauge invariance, it
is possible to show that the $\Delta {\cal L}_{i}$s have the form 
\[
\Delta {\cal L}_{i}=-\frac{1}{2}\Phi Q_{0}\widetilde{Q}_{i}Q_{0}\Phi =\frac{1%
}{2}\Phi Q_{0}\widetilde{Q}_{i}{\rm E}_{\Phi }+{\cal O}(\Phi ^{3}), 
\]
therefore they are proportional to the field equations, up to vertices. This
means that the $\Delta {\cal L}_{i}$s can be removed with field
redefinitions plus redefinitions of the vertex couplings, without affecting
the $\Phi $-propagator. Moreover, since the higher-derivative quadratic
counterterms depend on $Q_{0}\Phi $ and not just $\Phi $, the $\Delta {\cal L%
}_{i}$s can be promoted to gauge-invariant expressions, with additions of
vertices. This ensures that the field redefinitions are BRS\ covariant.

The meaning of the property that we have just illustrated is a sort of
``uniqueness'' of the propagator.

In practice, the reason why the quadratic divergent terms are not dangerous
is that they are always BRS-exact (because proportional to the field
equations) up to vertices. Below, I use these guidelights to prove the
theorem in the case of spin 3/2 fields. To generalize the argument when the
fields are expanded around non-trivial vacuum configurations, the vacua have
to be sufficiently ``nice'', for example constant, as in the case of scalar
vacuum expectation values, or with a constant curvature, as in the case of
gravity. On a background with constant curvature the curvature tensors can
be expressed in terms of the metric tensor, and the classification of the
counterterms becomes relatively simple. Then the proof of the theorem is
more or less straightforward. This is shown explicitly in section 7.

Finally, expanding the counterterms around other non-trivial configurations,
such as instantons or a space-time with non-constant curvature, the theorem
generalizes in the sense that it restricts the form of the lagrangian, but
the propagator contains quadratic terms with an arbitrary number of
derivatives.

\bigskip

In the next section, I study the most general quantum field theories of
fields of spin 0, 1/2, 1 in arbitrary dimensions. In section 5 I study the
fields of spin 3/2. The fields of spin 0 and 1/2 can be massive, the fields
of spin 1 and 3/2 are assumed to be massless. Gravity without a cosmological
constant is studied in the section 6. The theorem is generalized to theories
with a cosmological constant in section 7.

I work in the Euclidean framework. In flat space, all space-time indices are
low. By convention, I take $\kappa $ to have the universal dimension $-1,$
in units of mass. For simplicity, I also assume in most cases that the gauge
theories are parity invariant. If they are not parity invariant,
Chern-Simons terms have to be added in odd dimensions.

\section{Fields of spin 0, 1/2 and 1}

I\ begin with the fields of spin 0 and 1/2, which are trivial. The spin-1
case is preparatory for gravity.

\medskip

{\bf Spin 0.} The action reads 
\begin{equation}
S[\varphi ]=\frac{1}{2}\int \varphi \left( -\Box +m^{2}\right) \varphi +%
{\cal O}(3){\rm ,}  \label{ac}
\end{equation}
where ${\cal O}(3)$ denote terms that are cubic in the fluctuations around
the vacuum configuration. The field equations have the form ${\rm E}%
_{\varphi }=0$, where 
\[
{\rm E}_{\varphi }\equiv \left( -\Box +m^{2}\right) \varphi +{\cal O}(2). 
\]
If the theory contains other fields than $\varphi $ the terms ${\cal O}(2)$
can contain them.

The divergent terms quadratic in $\varphi $ have the form 
\begin{eqnarray}
\sum_{p=1}^{\infty }\kappa ^{2p}\sum_{j=0}^{p+1}a_{j}\ m^{2j}\ \varphi
\left( -\Box +m^{2}\right) ^{p+1-j}\varphi &=&\sum_{p=1}^{\infty }\kappa
^{2p}\sum_{j=0}^{p}a_{j}\ m^{2j}\ \varphi \left( -\Box +m^{2}\right) ^{p-j}%
{\rm E}_{\varphi }+  \label{quadra} \\
&&+\sum_{p=1}^{\infty }\kappa ^{2p}a_{p+1}\ m^{2p+2}\ \varphi ^{2}+{\cal O}%
(3),  \nonumber
\end{eqnarray}
where $a_{j}$ are certain divergent coefficients. A field redefinition 
\begin{equation}
\varphi \rightarrow \widetilde{\varphi }(\varphi )=\varphi
-\sum_{p=1}^{\infty }\kappa ^{2p}\sum_{j=0}^{p}a_{j}\ m^{2j}\ \left( -\Box
+m^{2}\right) ^{p-j}\varphi ,  \label{redefa}
\end{equation}
removes, up to vertices, the quadratic divergent terms of (\ref{quadra}),
but 
\begin{equation}
\sum_{p=1}^{\infty }\kappa ^{2p}a_{p+1}\ m^{2(p+1)}\ \varphi ^{2}.
\label{fin}
\end{equation}
This term can be reabsorbed with a renormalization of the mass.

The argument generalizes immediately to the case when the scalars are
coupled to gauge fields, because the covariant derivatives can be commuted
up to vertex terms. This is true also for the coupling to gravity, if the
vacuum metric is flat. This is the reason why we assume, for the moment,
that the theory has no cosmological constant. Note that in the presence of
gravity the scalar fields have to be massless, otherwise a cosmological
constant is induced by renormalization and the vacuum metric cannot be flat.
Under these assumptions, the quadratic divergent terms can be reabsorbed
with a gauge-covariant field redefinition, an immediate generalization of (%
\ref{redefa}). By the same argument, the ${\cal O}(3)$ terms of (\ref{ac})
need not contain $\left( -\Box +m^{2}\right) \varphi $.

\medskip

{\bf Spin 1/2.} In the case of the fermion, the lagrangian is 
\[
{\cal L}=\overline{\psi }\left( D\!\!\!\!\slash+m\right) \psi +{\cal O}(3). 
\]
The field equations are ${\rm E}_{\psi }=0$, where 
\[
{\rm E}_{\psi }\equiv \left( D\!\!\!\!\slash+m\right) \psi +{\cal O}(2). 
\]
We can have only quadratic counterterms of the form 
\[
I_{n,k}=\overline{\psi }D_{\lambda _{1}}\cdots D_{\lambda _{n}}\gamma _{\nu
_{1}}\cdots \gamma _{\nu _{k}}\psi ,\qquad n>1, 
\]
with variously contracted indices. Proceeding by induction in $k$, we can
assume that no $\nu _{i}$s are contracted together, otherwise we reduce to a
case with a lower $k$. Therefore, all $\nu _{i}$s are contracted with the $%
\lambda _{i}$s and possibly some $\lambda _{i}$s are contracted together.
Commuting the covariant derivatives, we get, up to vertex terms and terms
with a lower $k$, expressions of the form $\overline{\psi }D\!\!\!\!\slash%
^{q}\psi $, $q>1$. At the $\kappa ^{p}$th order the quadratic divergent
terms have the form 
\[
\kappa ^{p}\sum_{j=0}^{p+1}b_{j}\ m^{j}\ \overline{\psi }\left( D\!\!\!\!%
\slash+m\right) ^{p+1-j}\psi =\kappa ^{p}b_{p+1}\ m^{p+1}\ \overline{\psi }%
\psi +\kappa ^{p}\sum_{j=0}^{p}b_{j}\ m^{j}\ \overline{\psi }\left( D\!\!\!\!%
\slash+m\right) ^{p-j}{\rm E}_{\psi }+{\cal O}(3), 
\]
which can be reabsorbed with the gauge-covariant field redefinition 
\[
\psi \rightarrow \psi -\frac{1}{2}\sum_{p=1}^{\infty }\kappa
^{p}\sum_{j=0}^{p}b_{j}\ m^{j}\ \overline{\psi }\left( D\!\!\!\!\slash%
+m\right) ^{p-j}\psi , 
\]
up to vertices, plus a mass renormalization. By the same argument, the $%
{\cal O}(3)$ terms of the lagrangian need not contain $\left( D\!\!\!\!\slash%
+m\right) \psi $ and $\overline{\psi }\left( \overleftarrow{D\!\!\!\!\slash}%
+m\right) $.

\medskip

{\bf Spin 1}. Consider pure Yang-Mills theory in arbitrary dimension $d$
with lagrangian

\begin{equation}
{\cal L}=\frac{1}{\kappa ^{d-4}}\left[ \ \frac{F_{\mu \nu }^{2}}{4\alpha }%
+\sum_{n=1}^{\infty }\lambda _{n}\kappa ^{2n}\Im _{n}[F,\nabla ]\right] ,
\label{theory}
\end{equation}
where $\Im _{n}[F,\nabla ]$ denote collectively the gauge-invariant terms of
dimension $2n+4$ in units of mass, constructed with the field strength $F$
and the covariant derivative $\nabla $, with the condition that they are at
least cubic in $F$, up to total derivatives.

The field equations have the form 
\begin{equation}
\nabla _{\alpha }F_{\alpha \beta }={\cal O}(F^{2}).  \label{o2}
\end{equation}
The quadratic counterterms have the form 
\begin{equation}
F_{\mu \nu }\nabla _{\lambda _{1}}\cdots \nabla _{\lambda _{2n}}F_{\alpha
\beta }  \label{boo}
\end{equation}
with variously contracted indices. The derivatives can be interchanged up to
cubic terms. If an index $\lambda $ is contracted with an index of a field
strength $F$ in (\ref{boo}), such as in 
\begin{equation}
\int F_{\mu \nu }\nabla _{\lambda _{1}}\cdots \nabla _{\lambda }\cdots
\nabla _{\lambda _{2n}}F_{\lambda \beta }  \label{iop}
\end{equation}
we first move the covariant derivative $\nabla _{\lambda }$ till it acts
directly on $F_{\lambda \beta }$, then rewrite the counterterm as the sum of
a term proportional to the field equations (\ref{o2}) plus cubic terms.

If no index $\lambda $ of (\ref{boo}) is contracted with the indices of the
field strengths, then the counterterm has the form

\begin{equation}
\int F_{\mu \nu }\Box ^{p}F_{\mu \nu }  \label{rrboo}
\end{equation}
with $p=0,1,2,\ldots $ For $p=0$ the counterterm is removed with a
renormalization of the gauge coupling $\alpha $. For $p>0$ we rewrite (\ref
{rrboo}), up to vertex counterterms, in the form 
\[
\int ~\nabla _{\alpha }F_{\mu \nu }\Box ^{p-1}\nabla _{\alpha }F_{\mu \nu } 
\]
and use the Bianchi identity 
\begin{equation}
\nabla _{\alpha }F_{\mu \nu }+\nabla _{\nu }F_{\alpha \mu }+\nabla _{\mu
}F_{\nu \alpha }=0,  \label{biancoboo}
\end{equation}
to go back to the case (\ref{iop}).

The terms proportional to the field equations can be reabsorbed with a BRS\
covariant field redefinition of the vector potential $A_{\mu },$ of the form 
\begin{eqnarray*}
A_{\mu }\rightarrow A_{\mu } &&+c_{1}\alpha \nabla ^{\nu }F_{\mu \nu
}+\alpha ^{2}\left[ c_{2}F_{\mu \alpha }\nabla _{\nu }F^{\nu \alpha
}+c_{3}F^{\nu \alpha }\nabla _{\nu }F_{\mu \alpha }\right] \\
&&+\alpha ^{2}\left[ c_{4}\nabla _{\mu }F^{2}+c_{5}\nabla ^{\alpha }\nabla
_{\alpha }\nabla ^{\nu }F_{\mu \nu }\right] +{\cal O}(\alpha ^{3}).
\end{eqnarray*}
Here the $c_{i}$ are numerical coefficients.

In conclusion, the lagrangian (\ref{theory}) preserves its form under
renormalization. I will say that (\ref{theory}) is ``renormalizable'', with
which I do not mean that it is predictive or renormalizable by power
counting, but simply that the missing terms (the higher-derivative terms
quadratic in $A_{\mu }$) are not generated by renormalization.

\section{Fields of spin 3/2}

In this section I\ treat the case of spin-3/2 fields, first in four
dimensions and then in arbitrary dimension greater than two. I prove that
every gauge-invariant quadratic counterterm is necessarily proportional to
the field equations up to vertices.

\medskip

{\bf Spin 3/2 in four dimensions.} The lagrangian reads 
\[
{\cal L}=i\overline{\psi }_{\mu }\gamma _{5}\varepsilon _{\mu \nu \rho
\sigma }\partial _{\rho }\gamma _{\sigma }\psi _{\nu }+{\rm vertices.} 
\]
In momentum space, let us define the matrix 
\[
M_{\mu \nu }(k)=\gamma _{5}\varepsilon _{\mu \nu \rho \sigma }k_{\rho
}\gamma _{\sigma }. 
\]
The field equations read $M_{\mu \nu }(k)\psi _{\nu }(k)+{\cal O}(2)=0$.
Gauge invariance to the lowest order in the quantum fluctuations around the
vacuum $\psi _{\mu }=0$ ($\delta \psi _{\mu }=\partial _{\mu }\epsilon $) is
expressed by 
\begin{equation}
k_{\mu }M_{\mu \nu }(k)=0,\qquad M_{\mu \nu }(k)k_{\nu }=0.  \label{gauge}
\end{equation}
Straightforward calculations prove that 
\begin{eqnarray}
M_{\mu \alpha }(k)\left( \delta _{\alpha \beta }-\frac{1}{4}\gamma _{\alpha
}\gamma _{\beta }\right) M_{\beta \nu }(k) &=&k^{2}\delta _{\mu \nu }-k_{\mu
}k_{\nu },  \nonumber \\
-2M_{\mu \alpha }(k)\left( \delta _{\alpha \beta }-\frac{1}{2}\gamma
_{\alpha }\gamma _{\beta }\right) M_{\beta \nu }(k) &=&k^{2}\gamma _{\mu \nu
}-\gamma _{\mu \alpha }k_{\alpha }k_{\nu }-k_{\mu }k_{\alpha }\gamma
_{\alpha \nu },  \label{rela} \\
M_{\mu \alpha }(k)\left( \delta _{\alpha \beta }-\frac{1}{4}\gamma _{\alpha
}\gamma _{\beta }\right) M_{\beta \gamma }(k)M_{\gamma \nu }(k)
&=&k^{2}M_{\mu \nu }(k),  \nonumber \\
M_{\mu \alpha }(k)k\!\!\!\slash M_{\alpha \nu }(k) &=&2k\!\!\!\slash\left(
k^{2}\delta _{\mu \nu }-k_{\mu }k_{\nu }\right) -k^{2}M_{\mu \nu }(k), 
\nonumber
\end{eqnarray}
where $\gamma _{\mu \nu }=[\gamma _{\mu },\gamma _{\nu }]$.

Now, let us consider the structure of the most general quadratic divergent
term. We need to distinguish the cases in which the power of $\kappa $ is
even or odd.

If the power of $\kappa $ is odd, we have a structure 
\begin{eqnarray}
\kappa ^{2p+1}\overline{\psi }_{\mu }(-k)\left( k^{2}\right) ^{p} &&\left[
\left( a+b\gamma _{5}\right) \left( k^{2}\delta _{\mu \nu }-k_{\mu }k_{\nu
}\right) +\right.  \nonumber \\
&&\left. +\left( c+d\gamma _{5}\right) \left( k^{2}\gamma _{\mu \nu }-\gamma
_{\mu \alpha }k_{\alpha }k_{\nu }-k_{\mu }k_{\alpha }\gamma _{\alpha \nu
}\right) \right] \psi _{\nu }(k),  \label{c1}
\end{eqnarray}
with $p\geq 0$. This expression is fixed imposing gauge invariance on the
most general element of the Clifford algebra, namely a linear combination of 
$1$, $\gamma _{5}$, $\gamma _{\mu }$, $\gamma _{\mu }\gamma _{5}$, $\gamma
_{\mu \nu }$. Relations (\ref{rela}) show that (\ref{c1}) is always
proportional to the matrix $M_{\mu \nu }(k)$, therefore to the field
equations up to vertices.

If the power of $\kappa $ is even, the most general gauge-invariant
structure is 
\begin{equation}
\kappa ^{2p}\overline{\psi }_{\mu }(-k)\left( k^{2}\right) ^{p-1}\left[
\left( a+b\gamma _{5}\right) k^{2}M_{\mu \nu }(k)+\left( c+d\gamma
_{5}\right) k\!\!\!\slash\left( k^{2}\delta _{\mu \nu }-k_{\mu }k_{\nu
}\right) \right] \psi _{\nu }(k),  \label{c2}
\end{equation}
with $p\geq 1$, and, using (\ref{rela}) again, it is proportional to the
matrix $M_{\mu \nu }(k)$, therefore to the field equations up to vertices.

We have just studied gauge-invariance to the lowest order in the expansion
around the vacuum, and shown that the quadratic divergent terms are
proportional to the free-field equations, up to vertices. This is not
enough. We have also to be sure that the quadratic divergent terms can be
extended to gauge-invariant expressions. Using (\ref{rela}) we see that both
(\ref{c1}) and (\ref{c2}) can be rewritten in the form 
\[
\overline{\psi }_{\mu }(-k)M_{\mu \alpha }(k)I_{\alpha \beta }(k)M_{\beta
\nu }(k)\psi _{\nu }(k), 
\]
for a suitable Lorentz matrix $I_{\alpha \beta }(k)$, polynomial in $k.$
Having factorized one matrix $M_{\mu \nu }$ to the right and one to the
left, we can also write these objects in the form 
\begin{equation}
\overline{W}_{\mu \alpha }(-k)J_{\alpha \beta }(k)W_{\beta \nu }(k),
\label{tbcov}
\end{equation}
with another Lorentz matrix $J_{\alpha \beta }(k)$, polynomial in $k.$ Here $%
\overline{W}_{\mu \nu }$ and $W_{\mu \nu }$ are the field strengths of the
fields, $\overline{W}_{\mu \nu }=\partial _{\mu }\overline{\psi }_{\nu }-$ $%
\partial _{\nu }\overline{\psi }_{\mu }$ and $W_{\mu \nu }=\partial _{\mu
}\psi _{\nu }-$ $\partial _{\nu }\psi _{\mu }$. Finally, (\ref{tbcov})
proves that in the higher-derivative quadratic divergent terms the fields $%
\overline{\psi }_{\mu }$ and $\psi _{\nu }$ never appear separately from
their field strengths. At this point, expression (\ref{tbcov}) can be easily
extended to a supergravity scalar.

\medskip

{\bf Spin 3/2 in higher dimensions. }The lagrangian reads 
\[
{\cal L}=i\overline{\psi }_{\mu }\gamma _{\mu \rho \nu }\partial _{\rho
}\psi _{\nu }+{\rm vertices,} 
\]
where $\gamma _{\mu \rho \nu }$ denotes the completely antisymmetrized
product of gamma matrices, $\gamma _{\mu \rho \nu }=\gamma _{\mu }\gamma
_{\rho }\gamma _{\nu }/6$ $+$ antisymmetrizations. In momentum space, let us
define the matrix 
\[
M_{\mu \nu }(k)=\gamma _{\mu \rho \nu }k_{\rho }. 
\]
Gauge invariance to the lowest order is expressed again by (\ref{gauge}).
The higher-dimensional generalizations of (\ref{rela}) read 
\begin{eqnarray}
\frac{1}{36}M_{\mu \alpha }(k)\left( \delta _{\alpha \beta }-\frac{d-3}{%
\left( d-2\right) ^{2}}\gamma _{\alpha }\gamma _{\beta }\right) M_{\beta \nu
}(k) &=&k^{2}\delta _{\mu \nu }-k_{\mu }k_{\nu },  \nonumber \\
-\frac{1}{18}M_{\mu \alpha }(k)\left( \delta _{\alpha \beta }-\frac{1}{d-2}%
\gamma _{\alpha }\gamma _{\beta }\right) M_{\beta \nu }(k) &=&k^{2}\gamma
_{\mu \nu }-\gamma _{\mu \alpha }k_{\alpha }k_{\nu }-k_{\mu }k_{\alpha
}\gamma _{\alpha \nu },  \label{rela2} \\
\frac{1}{36}M_{\mu \alpha }(k)\left( \delta _{\alpha \beta }-\frac{d-3}{%
\left( d-2\right) ^{2}}\gamma _{\alpha }\gamma _{\beta }\right) M_{\beta
\gamma }(k)M_{\gamma \nu }(k) &=&k^{2}M_{\mu \nu }(k),  \nonumber \\
M_{\mu \alpha }(k)k\!\!\!\slash M_{\alpha \nu }(k) &=&(d-2)k\!\!\!\slash%
\left( k^{2}\delta _{\mu \nu }-k_{\mu }k_{\nu }\right) -(d-3)k^{2}M_{\mu \nu
}(k).  \nonumber
\end{eqnarray}
A basis for the Clifford algebra is $\gamma _{\mu _{1}\cdots \mu _{i}}$ for $%
i=0,\cdots ,d$, where $d$ is the space-time dimension. The structure of the
most general quadratic divergent term is 
\[
\kappa ^{j-1}\overline{\psi }_{\mu }(-k)\gamma _{\mu _{1}\cdots \mu
_{i}}k_{\alpha _{1}}\cdots k_{\alpha _{j}}\psi _{\nu }(k),\qquad \kappa
^{j-1}\overline{\psi }_{\mu }(-k)\gamma _{\mu _{1}\cdots \mu _{i}}k_{\alpha
_{1}}\cdots k_{\alpha _{j}}\varepsilon _{\nu _{1}\cdots \nu _{n}}\psi _{\nu
}(k), 
\]
with variously contracted indices. Gauge invariance has not been imposed,
yet.

The indices of $\gamma _{\mu _{1}\cdots \mu _{i}}$ cannot be contracted
among themselves, by antisymmetry, nor with more than one $k$. In the first
class of terms (no $\varepsilon $ tensor), we easily find, after imposing
gauge invariance, the same terms as in (\ref{c1}) and (\ref{c2}), with no $%
\gamma _{5}$. To study the second class of terms (one $\varepsilon $
tensor), we use 
\[
\gamma _{\mu _{1}\cdots \mu _{i}}=\left( -1\right) ^{(n-i)(n-i+1)/2}\frac{1}{%
(n-i)!}\varepsilon _{\mu _{1}\cdots \mu _{i}\alpha _{i+1}\cdots \alpha
_{n}}\gamma _{\alpha _{i+1}\cdots \alpha _{n}}\gamma _{5}. 
\]
We get two $\varepsilon $-tensors and a $\gamma _{5}=\frac{1}{n!}\varepsilon
_{\mu _{1}\cdots \mu _{n}}\gamma _{\mu _{1}\cdots \mu _{n}}$. Replacing the
two $\varepsilon $s with a product of Kronecker deltas, we get the same
class of terms (\ref{c1}) and (\ref{c2}), this time multiplied by $\gamma
_{5}$ (equal to unity in odd dimensions). From this point on, the discussion
proceeds as in the four-dimensional case.

\section{Quantum gravity without a cosmological term}

In this section I discuss pure quantum gravity without a cosmological term.
The metric tensor is expanded around flat space, $g_{\mu \nu }=\delta _{\mu
\nu }+\phi _{\mu \nu }$. For the time being, I use the
dimensional-regularization technique, which is sensitive to the logarithmic
divergences, but ignores the power-like divergences (linear, quadratic,
etc.). More general regularization techniques will be discussed later. The $%
l^{{\rm th}}$-loop counterterms in $d$ dimensions are monomials of dimension 
\begin{equation}
2+l(d-2),  \label{mn}
\end{equation}
constructed with the covariant derivatives of the curvature tensor,
multiplied by $\kappa ^{(l-1)(d-2)}$.

I want to prove the renormalizability of the theory with lagrangian 
\begin{equation}
{\cal L}=\frac{1}{\kappa ^{d-2}}\sqrt{g}\left[ -\ R+\sum_{n=1}^{\infty
}\lambda _{n}\kappa ^{n(d-2)}\Im _{n}[R,\nabla ]\right] .  \label{lagrabo}
\end{equation}
Here $\Im _{n}[R,\nabla ]$ collectively denote the gauge invariant terms of
dimension $n(d-2)+2$ that can be constructed with three or more Riemann
tensors $R_{\mu \nu \rho \sigma }$ and the covariant derivative $\nabla $,
up to total derivatives. For example, $\Im _{1}[R,\nabla ]$ in six
dimensions is a linear combination of terms of the form $R_{\mu \nu \rho
\sigma }R_{\alpha \beta \gamma \delta }R_{\varepsilon \zeta \eta \xi },$
with all possible contractions of indices, but does not contain the terms $%
R_{\mu \nu \rho \sigma }\nabla _{\alpha }\nabla _{\beta }R_{\varepsilon
\zeta \eta \xi },$ which would affect the graviton propagator with higher
derivatives. If $d$ is odd only the even $n$s contribute to (\ref{lagrabo}).
The Ricci tensor $R_{\mu \nu }$ and the scalar curvature $R$ need not appear
explicitly in $\Im _{n}[R,\nabla ]$, since they can be removed by means of
field redefinitions.

The field equations of (\ref{lagrabo}) have the form E$_{\mu \nu }=0$, with 
\begin{equation}
{\rm E}_{\mu \nu }=R_{\mu \nu }-\frac{1}{2}g_{\mu \nu }R+{\cal O}\left(
R^{2}\right) ,\qquad  \label{flateq}
\end{equation}
so that 
\begin{equation}
R_{\mu \nu }={\rm E}_{\mu \nu }-\frac{1}{d-2}g_{\mu \nu }g^{\alpha \beta }%
{\rm E}_{\alpha \beta }+{\cal O}\left( R^{2}\right) ,\qquad R=-\frac{2}{d-2}%
g^{\mu \nu }{\rm E}_{\mu \nu }+{\cal O}\left( R^{2}\right) .  \label{flateq2}
\end{equation}
We have to prove that the quadratic counterterms are proportional to the
Ricci tensor or the scalar curvature, so that using (\ref{flateq}) they can
be traded for terms proportional to ${\rm E}_{\mu \nu }$, which can be
removed by means of covariant field redefinitions, plus terms cubic in the
curvature tensors, which can be removed renormalizing the couplings $\lambda
_{n}$ in (\ref{lagrabo}).

The quadratic counterterms that do not trivially contain the Ricci tensor
and the Ricci curvature have the form 
\begin{equation}
\int \sqrt{g}~R_{\alpha \beta \gamma \delta }\nabla _{\lambda _{1}}\cdots
\nabla _{\lambda _{2n}}R_{\mu \nu \lambda \rho },  \label{rr}
\end{equation}
with variously contracted indices. The derivatives can be freely
interchanged, because the difference between two terms (\ref{rr}) with
interchanged derivatives is a vertex counterterm. If an index $\lambda $ is
contracted with an index of a Riemann tensor $R_{\mu \nu \lambda \rho }$ in (%
\ref{rr}), 
\begin{equation}
\int \sqrt{g}~R_{\alpha \beta \gamma \delta }\nabla _{\lambda _{1}}\cdots
\nabla ^{\lambda }\cdots \nabla _{\lambda _{2n}}R_{\mu \nu \lambda \rho }
\label{rrp}
\end{equation}
we move the covariant derivative $\nabla _{\lambda }$ till it acts directly
on $R_{\mu \nu \lambda \rho }$ and then use the contracted Bianchi identity 
\begin{equation}
\nabla ^{\lambda }R_{\mu \nu \lambda \rho }=\nabla _{\mu }R_{\nu \rho
}-\nabla _{\nu }R_{\mu \rho }.  \label{cobianco}
\end{equation}
This produces terms proportional to ${\rm E}_{\mu \nu }$, plus cubic terms
in the curvature tensors. It remains to consider the quadratic counterterms
of the form 
\begin{equation}
\int \sqrt{g}~R_{\mu \nu \rho \sigma }\Box ^{p}R^{\mu \nu \rho \sigma }
\label{rr1}
\end{equation}
with $p=0,1,2,\ldots $ Let us first take $p>0$. Up to vertex counterterms,
we can replace (\ref{rr1}) by 
\begin{equation}
\int \sqrt{g}~\nabla _{\alpha }R_{\mu \nu \rho \sigma }\Box ^{p-1}\nabla
^{\alpha }R^{\mu \nu \rho \sigma }.  \label{rr2}
\end{equation}
Using the uncontracted Bianchi identity 
\begin{equation}
\nabla _{\alpha }R_{\mu \nu \rho \sigma }+\nabla _{\nu }R_{\alpha \mu \rho
\sigma }+\nabla _{\mu }R_{\nu \alpha \rho \sigma }=0,  \label{bianco}
\end{equation}
we go back to the case (\ref{rrp}).

We now consider the case $p=0$ in (\ref{rr1}). From (\ref{mn}) we see that
this case is relevant only in four ($d=4$, $l=1$) and three ($d=3$, $l=2$)
dimensions. In three dimensions the Riemann tensor is proportional to the
Ricci tensor, since the Weyl tensor is identically zero. In four dimensions
the combination 
\begin{equation}
\int \sqrt{g}{\rm G}=\int \sqrt{g}~\left( R_{\mu \nu \rho \sigma }R^{\mu \nu
\rho \sigma }-4R_{\mu \nu }R^{\mu \nu }+R^{2}\right)  \label{combina}
\end{equation}
is proportional to the Euler characteristic, which is zero at the
perturbative level.

We conclude that in arbitrary space-time dimension greater than two every
quadratic counterterm is proportional to the Ricci tensor or the Ricci
curvature, up to total derivatives and vertex counterterms, and can be
removed with a covariant field redefinition of the form 
\[
g_{\mu \nu }\rightarrow g_{\mu \nu }+\sum_{n=1}^{\infty }a_{n}\kappa
^{n(d-2)}\nabla ^{n(d-2)-2}R, 
\]
plus renormalizations of the couplings $\lambda _{n}$. Therefore the
lagrangian (\ref{lagrabo}) is renormalizable.

\bigskip

If we do not want to use the dimensional-regularization technique, but
prefer for example a conventional cut-off regularization, then the argument
can be generalized if the renormalized cosmological constant is still set to
zero. If there are no masses and parameters with positive dimension in units
of mass, this is always possible. If we do not want to set the renormalized
cosmological constant to zero, then we have to apply the results of the next
section.

In dimension $d>4$ the term (\ref{combina}) can appear among the
divergences, multiplied by a power of the cut-off $\overline{\Lambda }$: 
\[
\overline{\Lambda }^{d-4}f(\overline{\Lambda }\kappa ^{2})\sqrt{g}{\rm G} 
\]
This term is not present in (\ref{lagrabo}), but we can write a
renormalizable generalization of (\ref{lagrabo}), 
\begin{equation}
{\cal L}=\frac{1}{\kappa ^{d-2}}\sqrt{g}\left[ \ -R+\lambda \kappa ^{2}~{\rm %
G}+\sum_{n=1}^{\infty }\lambda _{n}^{\prime }\kappa ^{2n+2}\Im _{n}^{\prime
}[R,\nabla ]\right] ,  \label{lagrabu}
\end{equation}
suitable for a regularization technique that is sensitive to the powers of
the cut-off $\overline{\Lambda }$, in a subtraction scheme where the
renormalized cosmological constant vanishes. In (\ref{lagrabu}) $\Im
_{n}^{\prime }[R,\nabla ]$ denote the gauge-invariant terms of dimension $%
2n+4$ that can be constructed with three or more curvature tensors and the
covariant derivative, up to total derivatives. I have remarked in the
introduction that (\ref{combina}) is a vertex term in every space-time
dimension, since it is at least cubic in the quantum fluctuation $\phi _{\mu
\nu }$. This proves that the graviton propagator associated with the
lagrangian (\ref{lagrabu}) does not contain higher derivatives. In
particular, the field equations preserve the form (\ref{flateq}), since the
variation of $\int \sqrt{g}{\rm G}$ with respect to the metric is ${\cal O}%
(R^{2})$.

If there are no masses and parameters with positive dimension in units of
mass we can always choose a subtraction scheme where the renormalized
coupling constant $\lambda $ in front of ${\rm G}$ is set to zero. In this
case, we recover the results of the conventional dimensional-regularization
scheme.

Obviously, the renormalizability of (\ref{lagrabo}) and (\ref{lagrabu}) does
not depend on the expansion $g_{\mu \nu }=\delta _{\mu \nu }+\phi _{\mu \nu
} $. However, it is only with respect to this expansion that higher
derivatives do not appear in the graviton propagator. If, for example, we
expand the metric in (\ref{lagrabo}) or (\ref{lagrabu}) around an instanton
background, then the graviton propagator does contain higher derivatives.

\bigskip

The results proved so far generalize immediately to theories containing
massless fields of spins 0, 1/2, 1, 3/2 and 2 coupled together, since the
quadratic part of the action is just the sum of the quadratic parts of the
fields. The fields have to be massless in order to be allowed to set the
renormalized cosmological constant to zero.

\section{Quantum gravity with a cosmological term}

A cosmological term is always induced by renormalization when massive fields
are coupled to gravity, actually whenever the classical lagrangian contains
a dimensionful parameter with positive dimension in units of mass. Even
choosing a subtraction scheme where the quartic and quadratic divergences
are ignored by default, the masses are responsible for the appearance of
logarithmic divergences of the form 
\[
\frac{m^{d}}{\varepsilon ^{n}}f(m\kappa )\sqrt{g}, 
\]
which can be removed only with a redefinition of the cosmological constant $%
\Lambda $. It is therefore mandatory to generalize the theorem of the
previous sections to a non-vanishing $\Lambda $.

I start from classical gravity in $d$ dimensions with lagrangian 
\begin{equation}
{\cal L}_{0}=-\frac{1}{\kappa ^{d-2}}\sqrt{g}\ \left( R-\Lambda \right) .
\label{agra}
\end{equation}

At the quantum level, the lagrangian ${\cal L}_{0}$ and its field equations 
\begin{equation}
R_{\mu \nu }-\frac{1}{2}g_{\mu \nu }R+\frac{\Lambda }{2}g_{\mu \nu }=0,
\label{feq}
\end{equation}
are modified by the additions of infinitely many terms, necessary to
renormalize the divergences. The generic gravitational counterterm reads 
\begin{equation}
\int \sqrt{g}\ \left( \nabla _{\mu _{1}}\cdots \nabla _{\mu _{p_{1}}}\left.
R^{\alpha _{1}}\right. _{\beta _{1}\gamma _{1}\delta _{1}}\right) \cdots \
\left( \nabla _{\mu _{1}}\cdots \nabla _{\mu _{p_{m}}}\left. R^{\alpha
_{m}}\right. _{\beta _{m}\gamma _{m}\delta _{m}}\right) \   \label{term}
\end{equation}
with variously contracted indices.

\bigskip

{\bf Choice of the background.} We have to choose an appropriate
gravitational vacuum $\overline{g}_{\mu \nu }$ to define the quantum
fluctuations $h_{\mu \nu }$, 
\[
g_{\mu \nu }=\overline{g}_{\mu \nu }+h_{\mu \nu }. 
\]
The expansion of the Riemann tensor is $\left. R^{\mu }\right. _{\nu \rho
\sigma }=\left. \overline{R}^{\mu }\right. _{\nu \rho \sigma }+\left.
R^{(1)\ \mu }\right. _{\nu \rho \sigma }+\left. R^{(2)\ \mu }\right. _{\nu
\rho \sigma }+{\cal O}(h^{3})$, with 
\begin{equation}
\left. R^{(1)\ \mu }\right. _{\nu \rho \sigma }=\frac{1}{2}\left( \overline{%
\nabla }_{\rho }\overline{\nabla }_{\sigma }h_{\nu }^{\mu }+\overline{\nabla 
}_{\rho }\overline{\nabla }_{\nu }h_{\sigma }^{\mu }-\overline{\nabla }%
_{\rho }\overline{\nabla }^{\mu }h_{\nu \sigma }-\overline{\nabla }_{\sigma }%
\overline{\nabla }_{\rho }h_{\nu }^{\mu }-\overline{\nabla }_{\sigma }%
\overline{\nabla }_{\nu }h_{\rho }^{\mu }+\overline{\nabla }_{\sigma }%
\overline{\nabla }^{\mu }h_{\nu \rho }\right) .  \label{bobo}
\end{equation}
We do not need the explicit expression of $\left. R^{(2)\ \mu }\right. _{\nu
\rho \sigma }$, which can be found in \cite{thooftveltman}.

I first choose a vacuum metric $\overline{g}_{\mu \nu }$ satisfying the
field equations (\ref{feq})\ of the lagrangian (\ref{agra}),

\begin{equation}
\overline{R}_{\mu \nu }=\frac{\Lambda }{d-2}\overline{g}_{\mu \nu },
\label{aut}
\end{equation}
and study the counterterms (\ref{term}) to the second order in $h$. Later I
show that the vacuum metric is ``stable'' under renormalization, namely the
additional couplings do not affect $\overline{g}_{\mu \nu }$, but at most
renormalize the values of the cosmological constant and the Newton constant.

Now, the counterterms (\ref{term}) contain contributions quadratic in $h$
with arbitrary higher derivatives, such as 
\begin{equation}
\overline{\nabla }^{m_{1}}\overline{R}\cdots \overline{\nabla }^{m_{n}}%
\overline{R}\ \overline{\nabla }^{p}h_{\mu \nu }\ \overline{\nabla }%
^{q}h_{\rho \sigma }.  \label{diff}
\end{equation}
If the Riemann tensor $\overline{R}_{\mu \nu \rho \sigma }$ does not have a
prescribed form, the propagator of the graviton contains arbitrarily many
higher derivatives.

To have control on the counterterms, we may assume that the spacetime
manifold admits a metric with constant curvature and choose boundary
conditions such that the unique solution to (\ref{aut}) satisfies also 
\begin{equation}
\overline{R}_{\mu \nu \rho \sigma }=\frac{\Lambda }{\left( d-1\right) \left(
d-2\right) }\left( \overline{g}_{\mu \rho }\overline{g}_{\nu \sigma }-%
\overline{g}\Sb \mu \sigma  \\  \endSb \overline{g}_{\nu \rho }\right) .
\label{riema}
\end{equation}
In this case, the quadratic terms (\ref{diff}) simplify enormously and we
can generalize the proof of the previous section. The goal is to show that
on the spaces with constant curvature the higher-derivative quadratic terms
can be removed by means of covariant field redefinitions and vertex
renormalizations.

The metric of a space of constant curvature can always be written in the
form 
\[
{\rm d}s^{2}=\frac{{\rm d}x_{a}\ {\rm d}x^{a}}{\left( 1+\frac{\Lambda }{%
4(d-1)(d-2)}x_{a}x^{a}\right) ^{2}}, 
\]
in a suitable coordinate frame. Here the indices are raised and lowered with
the metric $\breve{g}_{\mu \nu }={\rm diag}(\varepsilon _{1},\ldots
,\varepsilon _{d})$, and $\varepsilon _{i}=\pm 1$ as appropriate.

Moreover, a Riemannian space has constant curvature if and only if it
(locally) admits an a group $G_{r}$ of motions, with $r=d(d+1)/2$ and if and
only if it (locally) admits an isotropy group $H_{s}$ of $s=d(d-1)/2$
parameters at each point \cite{call}.

The restriction (\ref{riema}) allows us to work on a class of interesting
spaces, such as de Sitter ($\Lambda >0$) and anti de Sitter ($\Lambda <0$).
The spaces (\ref{riema}) are the most symmetric spaces with a cosmological
constant, natural candidates for the perturbative vacuum of quantum gravity
with a cosmological constant. They are precisely the conformally flat spaces
satisfying (\ref{aut}). Indeed, it is immediate to prove that

{\it a conformally flat space satisfying (\ref{aut}) has constant curvature,
i.e. it satisfies also (\ref{riema}); conversely, a space of constant
curvature is conformally flat and satisfies (\ref{aut}).}

I define 
\[
\widehat{R}_{\mu \nu \rho \sigma }=R_{\mu \nu \rho \sigma }-\frac{\Lambda }{%
\left( d-1\right) \left( d-2\right) }\left( g_{\mu \rho }g_{\nu \sigma }-g\Sb
\mu \sigma  \\  \endSb g_{\nu \rho }\right) , 
\]
which is more convenient to study the $h$-expansion, since $\widehat{R}_{\mu
\nu \rho \sigma }$ is ${\cal O}(h)$.

\bigskip

{\bf Inductive hypothesis and strategy of the proof.} I\ assume, by
inductive hypothesis, that the complete lagrangian ${\cal L}$ is such that
its ${\cal O}(h)$- and ${\cal O}(h^{2})$- contributions come only from $%
{\cal L}_{0}$. The most general expression satisfying these conditions is 
\begin{equation}
{\cal L}=\frac{1}{\kappa ^{d-2}}\sqrt{g}\left[ \ -R+\Lambda +\lambda \kappa
^{2}~\widehat{{\rm G}}+\sum_{n=1}^{\infty }\lambda _{n}\kappa ^{2n+2}\Im
_{n}[\widehat{R},\nabla ,\Lambda ]\right]  \label{agra2}
\end{equation}
and the goal is to show that this lagrangian is renormalizable.

The object $\Im _{n}[\widehat{R},\nabla ,\Lambda ]$ collectively denotes the
gauge-invariant terms of dimension $2n+4$ that can be constructed with three
or more tensors $\widehat{R}_{\mu \nu \rho \sigma }$, the covariant
derivative $\nabla $, and powers of the cosmological constant $\Lambda $, up
to total derivatives. So, for example, $\Im _{1}[\widehat{R},\nabla ,\Lambda
]$ is a linear combination of terms of the form $\widehat{R}_{\mu \nu \rho
\sigma }\widehat{R}_{\alpha \beta \gamma \delta }\widehat{R}_{\varepsilon
\zeta \eta \xi }$ with all possible contractions of indices, but does not
contain the terms $\widehat{R}_{\mu \nu \rho \sigma }\nabla _{\alpha }\nabla
_{\beta }\widehat{R}_{\varepsilon \zeta \eta \xi }$, which would affect the $%
h$-propagator with higher derivatives. The contracted tensor $\widehat{R}%
_{\mu \nu }$ and the scalar $\widehat{R}$ need not appear explicitly in $\Im
_{n}[\widehat{R},\nabla ,\Lambda ]$, because they can be removed by means of
field redefinitions.

The object $\widehat{{\rm G}}$ is an appropriate generalization of the ${\rm %
G}$ of (\ref{combina}) and reads 
\begin{eqnarray*}
\widehat{{\rm G}} &=&\widehat{R}_{\mu \nu \rho \sigma }\widehat{R}^{\mu \nu
\rho \sigma }-4\widehat{R}_{\mu \nu }\widehat{R}^{\mu \nu }+\widehat{R}^{2}+%
\frac{4(d-3)}{(d-1)(d-2)}\Lambda \left( R-\Lambda \right) \\
&=&R_{\mu \nu \rho \sigma }R^{\mu \nu \rho \sigma }-4R_{\mu \nu }R^{\mu \nu
}+R^{2}-\frac{(d-3)(d-4)}{(d-1)(d-2)}\Lambda \left( 2R-\Lambda \right) .
\end{eqnarray*}
$\widehat{{\rm G}}$ is constructed so that the spacetime integral of $\sqrt{g%
}\ \widehat{{\rm G}}$ does not contain $h$-quadratic terms with higher
derivatives, when it is expanded around the background (\ref{riema}).
Precisely, after a straightforward, but lengthy calculation, using (\ref
{bobo}), we find 
\begin{equation}
\int \sqrt{g}\ \widehat{{\rm G}}=\frac{8(d-3)}{(d-1)(d-2)^{2}}\Lambda
^{2}\int \sqrt{\overline{g}}+{\cal O}\left( h^{3}\right) .  \label{bibi}
\end{equation}
This identity is a nontrivial fact and the key ingredient to prove our
theorem.

The $2N$th truncated theory is the theory with lagrangian 
\[
{\cal L}_{2N}=\frac{1}{\kappa ^{d-2}}\sqrt{g}\left[ \ -R+\Lambda +\lambda
\kappa ^{2}~\widehat{{\rm G}}+\sum_{n=1}^{N-2}\lambda _{n}\kappa ^{2n+2}\Im
_{n}[\widehat{R},\nabla ,\Lambda ]\right] , 
\]
where only the powers $\left( E\kappa \right) ^{p}\left( \Lambda \kappa
^{2}\right) ^{q}$ with $p+2q\leq 2N$ are kept, $E$ denoting the reference
energy scale. It is easy to see that at each order of the $\hbar $ expansion
we need to consider only a finite number of Feynman diagrams in the
truncated theory. Indeed, the powers of $\Lambda $ are bounded by the very
same definition of the truncation and, on dimensional grounds, the terms of $%
\Im _{n}[\widehat{R},\nabla ,\Lambda ]$ can be multiplied only by a
polynomial in $\lambda _{n}\kappa ^{2n}$, with $n\leq N-2$.

\bigskip

{\bf Field equations.} Due to (\ref{bibi}), the terms of ${\cal L}$
quadratic in $h$ come only from ${\cal L}_{0}$, so the field equations of $%
{\cal L}$ are ${\rm E}_{\mu \nu }=0$ with 
\begin{eqnarray}
{\rm E}_{\mu \nu } &=&\kappa ^{2}\frac{\delta S[\overline{g}+h]}{\delta
h^{\mu \nu }}=-\frac{1}{2}\overline{\nabla }^{2}h_{\mu \nu }+\frac{1}{2}%
\overline{g}_{\mu \nu }\overline{\nabla }^{2}h-\frac{1}{2}\overline{g}_{\mu
\nu }\overline{\nabla }_{\alpha }\overline{\nabla }_{\beta }h^{\alpha \beta
}-\frac{1}{2}\overline{\nabla }_{\mu }\overline{\nabla }_{\nu }h  \nonumber
\\
&&+\frac{1}{2}\overline{\nabla }_{\mu }\overline{\nabla }^{\alpha }h_{\alpha
\nu }+\frac{1}{2}\overline{\nabla }_{\nu }\overline{\nabla }^{\alpha }h_{\mu
\alpha }+\frac{\Lambda }{(d-1)(d-2)}h_{\mu \nu }+\frac{\Lambda (d-3)}{%
2(d-1)(d-2)}\overline{g}_{\mu \nu }h+{\cal O}\left( h^{2}\right)  \nonumber
\\
&&\qquad =R_{\mu \nu }-\frac{1}{2}g_{\mu \nu }\left( R-\Lambda \right) +%
{\cal O}\left( h^{2}\right) =\widehat{R}_{\mu \nu }-\frac{1}{2}g_{\mu \nu }%
\widehat{R}+{\cal O}\left( h^{2}\right) .  \label{guru}
\end{eqnarray}
where $h=h_{\mu \nu }\overline{g}^{\mu \nu }$ and indices are lowered and
raised using $\overline{g}_{\mu \nu }$. We have 
\begin{equation}
\widehat{R}_{\mu \nu }={\rm E}_{\mu \nu }-\frac{1}{d-2}g_{\mu \nu }g^{\alpha
\beta }{\rm E}_{\alpha \beta }+{\cal O}\left( h^{2}\right) ,\qquad \widehat{R%
}=-\frac{2}{d-2}g^{\mu \nu }{\rm E}_{\mu \nu }+{\cal O}\left( h^{2}\right) .
\label{feq22}
\end{equation}
We see that under the assumption that the lagrangian has the form (\ref
{agra2}), the field equations contain no higher-derivative terms linear in $%
h $ and therefore the form of the $h$-propagator is the standard one.
Moreover, the field equations (\ref{feq22}) are trivially solved by $h=0$.
This means that the background $\overline{g}_{\mu \nu }$ is not affected by
the infinitely many couplings $\lambda ,\lambda _{n}$. The background metric
is sensitive only to the value of the cosmological constant, which is
renormalized by radiative corrections, but otherwise the form of the vacuum
metric is stable under renormalization.

To proceed with the proof of our theorem, we\ need a more refined expression
for the field equations than (\ref{guru}) and (\ref{feq22}), that is to say 
\begin{equation}
{\rm E}_{\mu \nu }=\widehat{R}_{\mu \nu }-\frac{1}{2}g_{\mu \nu }\widehat{R}+%
{\cal O}\left( \widehat{R}^{2}\right) ,  \label{emunu}
\end{equation}
so that 
\begin{equation}
\widehat{R}_{\mu \nu }={\rm E}_{\mu \nu }-\frac{1}{d-2}g_{\mu \nu }g^{\alpha
\beta }{\rm E}_{\alpha \beta }+{\cal O}\left( \widehat{R}^{2}\right) ,\qquad 
\widehat{R}=-\frac{2}{d-2}g^{\mu \nu }{\rm E}_{\mu \nu }+{\cal O}\left( 
\widehat{R}^{2}\right) .  \label{feq2}
\end{equation}
To prove (\ref{feq2}) I start from the most general expression of the field
equations, which is, symbolically, 
\begin{equation}
{\rm E}_{\mu \nu }=\widehat{R}_{\mu \nu }-\frac{1}{2}g_{\mu \nu }\widehat{R}%
+\sum_{k\geq 1}\sum_{\{p_{i}\}}\prod_{i=1}^{k}\left( \nabla ^{p_{i}}\widehat{%
R}_{i}\right) .  \label{j}
\end{equation}
The last term of this formula collects the contributions the terms $\widehat{%
{\rm G}}$ and $\Im _{n}[\widehat{R},\nabla ,\Lambda ]$ in (\ref{agra2}). The
terms of (\ref{j}) with $k>1$ are ${\cal O}\left( \widehat{R}^{2}\right) $
and we do not need to discuss them. The terms with $k=1$ can come only from $%
\widehat{{\rm G}}$ and have the form 
\begin{equation}
b\widehat{R}_{\mu \nu }+cg_{\mu \nu }\widehat{R}+a^{\prime }\nabla _{\mu
}\nabla _{\nu }\widehat{R}+b^{\prime }\Box \widehat{R}_{\mu \nu }+c^{\prime
}g_{\mu \nu }\Box \widehat{R}  \label{iuppi}
\end{equation}
up to terms with a higher number of tensors $\widehat{R}$. These can be
included in the terms of (\ref{j}) with $k>1$. Due to (\ref{guru}), we know
that (\ref{iuppi}) has to be ${\cal O}\left( h^{2}\right) $. Using (\ref
{bobo}), we have 
\begin{eqnarray*}
\widehat{R}_{\mu \nu } &=&-\frac{1}{2}\overline{\nabla }_{\mu }\overline{%
\nabla }_{\nu }h+\frac{1}{2}\overline{\nabla }_{\mu }\overline{\nabla }%
^{\alpha }h_{\alpha \nu }+\frac{1}{2}\overline{\nabla }_{\nu }\overline{%
\nabla }^{\alpha }h_{\mu \alpha }-\frac{1}{2}\overline{\nabla }^{2}h_{\mu
\nu }+\frac{\Lambda (h_{\mu \nu }-h\overline{g}_{\mu \nu })}{(d-1)(d-2)}+%
{\cal O}\left( h^{2}\right) , \\
\widehat{R} &=&-\overline{\nabla }^{2}h+\overline{\nabla }^{\alpha }%
\overline{\nabla }^{\beta }h_{\alpha \beta }-\frac{\Lambda }{d-2}h+{\cal O}%
\left( h^{2}\right) .
\end{eqnarray*}
We see that no combination (\ref{iuppi}) can be ${\cal O}\left( h^{2}\right) 
$ and therefore the terms with $k=1$ are absent in (\ref{j}) and the field
equations of (\ref{agra2}) have the form 
\begin{equation}
{\rm E}_{\mu \nu }=\widehat{R}_{\mu \nu }-\frac{1}{2}g_{\mu \nu }\widehat{R}%
+\sum_{k>1}\sum_{\{p_{i}\}}\prod_{i=1}^{k}\nabla ^{p_{i}}\widehat{R}_{i},
\label{equatia}
\end{equation}
wherefrom (\ref{emunu}) and (\ref{feq2}) follow.

The result is completely general, that is to say it does not depend on the
expansion $g_{\mu \nu }=\overline{g}_{\mu \nu }+h_{\mu \nu }$ around a
background with constant curvature, although we used this expansion to prove
(\ref{emunu}). Equation (\ref{emunu}) can be proved more directly
differentiating $\int \sqrt{g}\widehat{{\rm G}}$ with respect to the metric.

Now we analyse the most general counterterms (\ref{term}) and prove that all
of them can be reabsorbed with covariant redefinitions of the metric tensor,
of the form 
\begin{equation}
g_{\mu \nu }\rightarrow g_{\mu \nu }+{\cal O}\left( \widehat{R}\right)
,\qquad {\rm i.e.}\qquad h_{\mu \nu }\rightarrow h_{\mu \nu }+{\cal O}\left(
h\right) ,  \label{redefu}
\end{equation}
plus renormalizations of the cosmological constant, the Newton constant and
the parameters $\lambda ,\lambda _{n}$ appearing in (\ref{agra2}), thus
preserving the structure (\ref{agra2}) to all orders in perturbation theory.

\bigskip

{\bf Analysis of the counterterms.} It is convenient to rewrite the
counterterms in the basis $\widehat{R}_{\mu \nu \rho \sigma }$, 
\begin{equation}
\int \sqrt{g}\ \left( \nabla _{\mu _{1}}\cdots \nabla _{\mu _{p_{1}}}\left. 
\widehat{R}^{\alpha _{1}}\right. _{\beta _{1}\gamma _{1}\delta _{1}}\right)
\cdots \ \left( \nabla _{\mu _{1}}\cdots \nabla _{\mu _{p_{m}}}\left. 
\widehat{R}^{\alpha _{m}}\right. _{\beta _{m}\gamma _{m}\delta _{m}}\right)
\   \label{terma}
\end{equation}
The terms (\ref{terma}) with $m\geq 3$ are of the type $\Im _{n}[\widehat{R}%
,\nabla ,\Lambda ]$ and are renormalized by means of redefinitions of the
couplings $\lambda _{n}$, $n\geq 1$. We need to study only the terms with $%
m=0,1,2$: 
\begin{eqnarray*}
&&I_{1}=\int \sqrt{g}\ ,\qquad \qquad \qquad I_{2}=\int \sqrt{g}\ \nabla
_{\mu }\cdots \nabla _{\mu _{p}}\left. \widehat{R}^{\alpha }\right. _{\beta
\gamma \delta }, \\
&&\qquad I_{3}=\int \sqrt{g}\ \left. \widehat{R}^{\alpha _{1}}\right.
_{\beta _{1}\gamma _{1}\delta _{1}}\nabla _{\mu _{1}}\cdots \nabla _{\mu
_{p}}\left. \widehat{R}^{\alpha _{2}}\right. _{\beta _{2}\gamma _{2}\delta
_{2}}.
\end{eqnarray*}

\bigskip

{\bf Terms }$I_{1}${\bf \ and }$I_{2}$. The term $I_{1}$ is a
renormalization of the cosmological constant. The term $I_{2}$ vanishes for $%
p>0$. For $p=0$ all contractions of indices in $I_{2}$ give 
\begin{equation}
\int \sqrt{g}\ \widehat{R}=\int \sqrt{g}\ \left( R-\frac{d}{d-2}\Lambda
\right) ,  \label{i2}
\end{equation}
which can be reabsorbed with renormalizations of the Newton constant and the
cosmological constant.

\bigskip

{\bf Terms }$I_{3}${\bf \ containing }$\widehat{R}_{\mu \nu }${\bf \ and }$%
\widehat{R}${\bf .} Using (\ref{feq2}), the terms $I_{3}$ containing $%
\widehat{R}_{\mu \nu }$ and $\widehat{R}$ are proportional to the field
equations up to higher-order terms in $\widehat{R}$. We have objects of the
form 
\[
\int \sqrt{g}~\widehat{R}_{\alpha \beta \gamma \delta }\nabla _{\lambda
_{1}}\cdots \nabla _{\lambda _{2n}}\widehat{R}_{\mu \nu }=\int \sqrt{g}~%
\widehat{R}_{\alpha \beta \gamma \delta }\nabla _{\lambda _{1}}\cdots \nabla
_{\lambda _{2n}}{\rm E}_{\mu \nu }+{\cal O}\left( \widehat{R}^{3}\right) , 
\]
with all possible contractions of indices. Divergences of this form can be
reabsorbed with a redefinition of the metric tensor of the form 
\[
g_{\mu \nu }\rightarrow g_{\mu \nu }+\nabla _{\lambda _{1}}\cdots \nabla
_{\lambda _{2n}}\widehat{R}_{\alpha \beta \gamma \delta },\qquad {\rm i.e.}%
\qquad h_{\mu \nu }\rightarrow h_{\mu \nu }+\nabla _{\lambda _{1}}\cdots
\nabla _{\lambda _{2n}}\widehat{R}_{\alpha \beta \gamma \delta }, 
\]
plus renormalizations of the couplings $\lambda _{n}$, $n\geq 1$.

\bigskip

{\bf Terms }$I_{3}${\bf \ not containing }$\widehat{R}_{\mu \nu }${\bf \ and 
}$\widehat{R}${\bf .} The terms $I_{3}$ that do not trivially contain $%
\widehat{R}_{\mu \nu }$ and $\widehat{R}$ have the form 
\begin{equation}
\int \sqrt{g}~\widehat{R}_{\alpha \beta \gamma \delta }\nabla _{\lambda
_{1}}\cdots \nabla _{\lambda _{p}}\widehat{R}_{\mu \nu \lambda \rho },
\label{brix}
\end{equation}
with variously contracted indices.

$i$) The terms (\ref{brix}) with $p>0$ can be treated as the terms (\ref{rr}%
) studied in the absence of a cosmological constant. First, let us observe
that the Bianchi identity (\ref{bianco}) holds also with hatted tensors: 
\begin{equation}
\nabla _{\alpha }\widehat{R}_{\mu \nu \rho \sigma }+\nabla _{\nu }\widehat{R}%
_{\alpha \mu \rho \sigma }+\nabla _{\mu }\widehat{R}_{\nu \alpha \rho \sigma
}=0.  \label{biagio}
\end{equation}
When covariant derivatives are commuted in (\ref{brix}), we get terms with a
higher number of tensors $\widehat{R}$ plus terms of the form (\ref{brix})
with a lower value of $p$, but belonging to the same truncation $\Gamma
^{(N)}$ of the theory: 
\begin{eqnarray}
\left[ \nabla _{\alpha },\nabla _{\beta }\right] T_{\mu _{1}\cdots \mu _{n}}
&=&-\sum_{i=1}^{n}\left. \widehat{R}_{{}}^{\rho }\right. _{\mu _{i}\alpha
\beta }T_{\mu _{1}\cdots \mu _{i-1}\rho \mu _{i+1}\cdots \mu _{n}}
\label{bia} \\
&&-\frac{\Lambda }{(d-1)(d-2)}\sum_{i=1}^{n}\left( T_{\mu _{1}\cdots \mu
_{i-1}\alpha \mu _{i+1}\cdots \mu _{n}}g_{\mu _{i}\beta }-T_{\mu _{1}\cdots
\mu _{i-1}\beta \mu _{i+1}\cdots \mu _{n}}g_{\mu _{i}\alpha }\right) . 
\nonumber
\end{eqnarray}

If in (\ref{brix}) an index of the covariant derivatives is contracted with
an index of a tensor $\widehat{R}_{\mu \nu \lambda \rho }$, we have terms of
the form 
\begin{equation}
\int \sqrt{g}~\widehat{R}_{\alpha \beta \gamma \delta }\nabla _{\lambda
_{1}}\cdots \nabla ^{\lambda }\cdots \nabla _{\lambda _{p}}\widehat{R}_{\mu
\nu \lambda \rho }.  \label{brix2}
\end{equation}
Here we commute the covariant derivatives till $\nabla _{\lambda }$ acts
directly on $\widehat{R}_{\mu \nu \lambda \rho }$ and then use the Bianchi
identity, obtaining terms proportional to the field equations plus
higher-order terms in $\widehat{R}$, plus terms (\ref{brix}) with lower
values of $p$ belonging to the same truncation level.

If in (\ref{brix}) no index of the covariant derivatives is contracted with
an index of a tensor $\widehat{R}_{\mu \nu \lambda \rho }$, we have terms of
the form

\[
\int \sqrt{g}~\widehat{R}_{\mu \nu \rho \sigma }\Box ^{p}\widehat{R}^{\mu
\nu \rho \sigma } 
\]
with $p=1,2,\ldots $ these objects can be rewritten, after commuting the
derivatives a sufficient number of times, as 
\[
\int \sqrt{g}~\nabla _{\alpha }\widehat{R}_{\mu \nu \rho \sigma }\Box
^{p-1}\nabla ^{\alpha }\widehat{R}^{\mu \nu \rho \sigma }, 
\]
plus higher orders in $\widehat{R}$ and terms of the form (\ref{brix}) with
lower values of $p$. Using the Bianchi identity we go back to the case (\ref
{brix2}).

Proceeding inductively in $p$, we can lower the value of $p$ arbitrarily,
remaining in the same truncation. In the end, we need to discuss only the $%
p=0$ terms.

$ii$) The terms (\ref{brix}) with $p=0$ are

\[
\int \sqrt{g}~\widehat{R}_{\alpha \beta \gamma \delta }\widehat{R}_{\mu \nu
\lambda \rho }. 
\]
The possible contractions of indices give 
\begin{equation}
\int \sqrt{g}~\widehat{R}^{\mu \nu \lambda \rho }\widehat{R}_{\mu \nu
\lambda \rho },\qquad \int \sqrt{g}~\widehat{R}^{\mu \nu }\widehat{R}_{\mu
\nu },\qquad \int \sqrt{g}~\widehat{R}^{2}.  \label{list}
\end{equation}
The second and third term of this list are proportional to the field
equations E$_{\mu \nu }$ up to higher orders in $\widehat{R}$. The first
term of (\ref{list}) is $\int \sqrt{g}\widehat{{\rm G}}$ plus a term $I_{1}$%
, a term $I_{2}$, the second and third terms of (\ref{list}).

Concluding, up to higher orders in $\widehat{R}$, the first term of (\ref
{list}) can be removed with renormalizations of $\lambda $, the cosmological
constant and the Newton constant, plus a covariant redefinition of the
metric tensor.

\bigskip

We have therefore proven that the structure (\ref{agra2}) is preserved to
every order and the $h$-propagator does not contain higher derivatives. For
each truncation $\Gamma ^{(N)}$ of the theory the removal of the divergences
requires a finite number of steps at each order of the $\hbar $ expansion.

\bigskip

{\bf Background independence.} Going through the proof of the theorem, we
see the the key-ingredients, which are (\ref{biagio}), (\ref{bia}) and (\ref
{feq2}), do not depend on the metric $\overline{g}_{\mu \nu }$ around which
the expansion is performed. The most general version of our theorem is:

{\it the action} 
\begin{equation}
S=\frac{1}{\kappa ^{d-2}}\int \sqrt{g}\left[ \ -R+\Lambda +\lambda \kappa
^{2}~\widehat{{\rm G}}+\sum_{n=1}^{\infty }\lambda _{n}\kappa ^{2n+2}\Im
_{n}[\widehat{R},\nabla ,\Lambda ]\right]  \label{answer}
\end{equation}
{\it is renormalizable with redefinitions of the cosmological constant, the
Newton constant and the parameters }$\lambda ,\lambda _{n}${\it , plus a
covariant redefinition of the metric tensor.}

The choice of a vacuum metric $\overline{g}_{\mu \nu }$ with constant
curvature is relevant only to the form of the graviton propagator. If the
vacuum metric satisfies (\ref{riema}), the graviton propagator of (\ref
{answer}) does not contain higher derivatives. If the vacuum metric does not
satisfy (\ref{riema}), then the graviton propagator of (\ref{answer}) does
contain higher derivatives.

\bigskip

{\bf Coupling to matter.} Finally, the theorem generalizes straightforwardly
to the case when the fields of spin 0, 1/2, 1, 3/2 and 2 are coupled
together. It is sufficient to recall that by spin conservation the mixed
quadratic terms, e.g. the spin-1/spin-2 quadratic terms 
\begin{equation}
\int \sqrt{g}~F_{\alpha \beta }\nabla _{\lambda _{1}}\cdots \nabla _{\lambda
_{p}}\widehat{R}_{\mu \nu \lambda \rho }  \label{mix}
\end{equation}
are actually vertices up to terms proportional to the field equations. For
example, in the case of (\ref{mix}) the covariant derivatives can be
commuted up to vertices and terms with a lower value of $p$ belonging to the
same truncation of the theory. If the index of a covariant derivative is
contracted with an index of $F$ or $\widehat{R}$, then, commuting the
covariant derivatives a sufficient number of times and using the Bianchi
identities, we obtain terms proportional to the field equations plus
vertices. So, we can assume that the indices of the covariant derivatives
are contracted among themselves. This leads to objects of the form 
\[
\int \sqrt{g}~F_{\alpha \beta }\Box ^{p}\widehat{R}_{\mu \nu \lambda \rho }, 
\]
which admit no non-trivial contraction.

\section{Perturbative vacuum and the vacuum of quantum gravity}

In this section I\ make some illustrative comments on the perturbative and
quantum vacua of the theory defined by the lagrangian (\ref{answer}). A more
detailed analysis of perturbation theory will be presented in a separate
paper.

The metric of constant curvature is such that the graviton propagator is
two-derivative. If the space-time manifold admits a metric of constant
curvature, we choose boundary conditions such that the unique solution to
the Einstein equations $\widehat{R}_{\mu \upsilon }-g_{\mu \nu }\widehat{R}%
/2=0$ is precisely that metric. Then the metric of constant curvature is
also the unique solution to the complete field equations E$_{\mu \nu }=0$ of
the action (\ref{answer}), where E$_{\mu \nu }$ is given by (\ref{emunu})
and (\ref{equatia}). This means that the metric of constant curvature is an
extremal of (\ref{answer}).

It is reasonable to expect that the quantum corrections, which certainly
change the vacuum metric in the bulk of the space-time manifold, do not
affect the behavior of the metric at the boundary of the space-time
manifold, although the value of the asymptotic curvature gets renormalized.
So, if the perturbative-vacuum metric is of constant curvature, the vacuum
of quantum gravity, i.e. the minimum of the quantum action $\Gamma [g_{\mu
\nu }]$, has an asymptotically constant curvature.

However, the metric of constant curvature is not a minimum of the action (%
\ref{answer}). The second derivative of (\ref{answer}), calculated on the
vacuum metric $\overline{g}_{\mu \nu }$ of constant curvature, can be
written explicitly, because we have shown that the quadratic terms of (\ref
{answer}) receive contributions only from the Einstein and cosmological
terms. We have 
\begin{eqnarray*}
\frac{1}{2} &&\int h_{\mu \nu }\left. \frac{\delta ^{2}S}{\delta g_{\mu \nu
}\delta g_{\rho \sigma }}\right| _{g_{\alpha \beta }=\overline{g}_{\alpha
\beta }}h_{\rho \sigma }=\frac{1}{4\kappa ^{d-2}}\int \sqrt{\overline{g}}\ 
{\rm H}[\widetilde{h},\overline{g}], \\
&&{\rm H}[\widetilde{h},\overline{g}]=\left( \overline{\nabla }_{\mu }%
\widetilde{h}_{\nu \rho }\right) ^{2}-\frac{1}{d-2}\left( \overline{\nabla }%
_{\mu }\widetilde{h}\right) ^{2}-2\left( \overline{\nabla }^{\mu }\widetilde{%
h}_{\mu \nu }\right) ^{2}+\frac{2\Lambda }{(d-1)(d-2)}\left( \widetilde{h}%
_{\mu \nu }^{2}+\frac{1}{d-2}\widetilde{h}^{2}\right) ,
\end{eqnarray*}
where $\widetilde{h}_{\mu \nu }=h_{\mu \nu }-\overline{g}_{\mu \nu }h/2$. H$%
[h,\overline{g}]$ is not positive definite, due to the known
negative-definite contribution of the conformal factor. For example, $h_{\mu
\nu }=h\overline{g}_{\mu \nu }/d$ gives 
\[
{\rm H}[\widetilde{h},\overline{g}]=-4\frac{d-1}{d^{2}(d-2)}\left( \overline{%
\nabla }_{\mu }\widetilde{h}\right) ^{2}+2\Lambda \widetilde{h}^{2}\frac{d-1%
}{d^{2}(d-2)^{2}}. 
\]

This means that\ either the total action (\ref{answer}) is not bounded from
below (in the Euclidean framework) or, if it is bounded from below, its
absolute minimum is not the metric of constant curvature. In the latter case
different boundary conditions select the true perturbative vacuum and the
quantum vacuum need not have an asymptotically constant curvature.

On the other hand, the perturbative calculations (propagator, vertices and
Feynman diagrams) of the theory (\ref{answer}) expanded around a metric of
constant curvature appear to be well-defined, at least if the curvature is
negative (anti-de Sitter space) \cite{dan}. When the curvature is positive
the Green functions behave less nicely \cite{turyn}. In particular, the
perturbative calculations around anti-de Sitter space do not exhibit the
typical difficulties of the expansion around a ``wrong'' configuration. For
comparison, let us take a scalar-field theory with potential $V\left( \phi
\right) =-\mu ^{2}\phi ^{2}/2+g^{2}\phi ^{4}/4$. A perturbative expansion
around the local maximum $\phi =0$ is problematic, because the propagator is
tachyonic. In principle, the physical results should not depend on the
configuration $\overline{\phi }$ around which the expansion in performed,
but the boundary conditions are crucial to select the appropriate $\overline{%
\phi }$. Precisely, two expansions around configurations $\overline{\phi }%
_{1}$ and $\overline{\phi }_{2}$ satisfying the same boundary conditions are
equivalent (after appropriate resummations). Since the local maximum of $%
V\left( \phi \right) $ tends to zero at infinity, while the minimun tends to 
$v=\mu /g$, we have no reason to expect that the expansion around $\overline{%
\phi }=v$ is equivalent to the expansion around $\overline{\phi }=0$, even
admitting that we can make sense of the it (e.g. with an analytical
continuation from negative $\mu ^{2}$).

Thus, the best criterion to choose the perturbative vacuum in a
non-renormalizable theory is not necessarily the minimum of the classical
action. Ultimately, it is not even necessary that the classical action,
which has no direct physical significance, be bounded from below. The first
check is to see whether the perturbative expansion around the temptative
vacuum is well-defined or not. Therefore, the metric of negative constant
curvature is a good candidate for the right perturbative vacuum of quantum
gravity, even if it is not the absolute minimum of the action (\ref{answer}%
). Then, we are allowed to argue that vacuum of quantum gravity has a
negative asymptotically constant curvature.

\section{Conclusions}

The results of this paper suggest that the theories with infinitely many
couplings can be studied in a perturbative sense also at high energies.
Various questions are well-posed and can be answered. In particular, a whole
class of lagrangian terms is not turned on by renormalization, if it is
absent at the tree level. For example, the lagrangian of quantum gravity in
arbitrary space-time dimension greater than two can be conveniently
organized as the sum 
\begin{equation}
{\cal L}=\frac{1}{\kappa ^{d-2}}\sqrt{g}\left[ \ -R+\Lambda +\lambda \kappa
^{2}~\widehat{{\rm G}}+\sum_{n=1}^{\infty }\lambda _{n}\kappa ^{2n+2}\Im
_{n}[\nabla ,\widehat{R},\Lambda ]\right] ,  \label{boi}
\end{equation}
where $\widehat{{\rm G}}$ and $\Im _{n}[\nabla ,\widehat{R},\Lambda ]$ are
defined in the paper. The lagrangian (\ref{boi}) is ``renormalizable'',
namely preserves its form under renormalization.

If the theory has no cosmological constant or the space-time manifold admits
a metric of constant curvature, the propagators of the fields are not
affected by higher derivatives. The metric of constant curvature is an
extremal, but neither an absolute minimum, nor a local minimum of the action
(\ref{boi}). Therefore, either the action (\ref{boi}) is not bounded from
below or there exists a different perturbative minimum. Some considerations
suggest that the metric of negative constant curvature is the right
perturbative vacuum to formulate the perturbative expansion of the theory (%
\ref{boi}), even if it is not a minimum of the classical action. Then we can
argue that the quantum vacuum has a negative asymptotically constant
curvature.

It would be interesting to see if the theorem proven here can be further
generalized to treat the vertices containing higher derivatives of the
metric.

The results of this paper might also be useful in effective field theory,
for a more convenient treatment of the radiative corrections to the
low-energy limit, and to properly address the cosmological constant problem.

\vskip .6truecm {\bf Acknowledgements}

\vskip .3truecm

I thank M. Porrati, P. Menotti and M. Mintchev for discussions and D.Z.
Freedman for correspondence.


\begin{thebibliography}{99}
\bibitem{thooftveltman}  G. 't Hooft and M. Veltman, One-loop divergences in
the theory of gravitation, Ann. Inst. Poincar\`{e}, 20 (1974) 69.

\bibitem{wein}  S. Weinberg, Ultraviolet divergences in quantum theories of
gravitation, in {\it An Einstein centenary survey}, Edited by S. Hawking and
W. Israel, Cambridge University Press, Cambridge 1979.

\bibitem{stelle}  K.S. Stelle, Renormalization of higher derivative quantum
gravity, Phys. Rev. D 16 (1977) 953.

\bibitem{sagnotti}  M.H. Goroff and A. Sagnotti, The ultraviolet behavior of
Einstein gravity, Nucl. Phys. B 266 (1986) 709.

\bibitem{me}  D. Anselmi, Removal of divergences with the Batalin-Vilkovisky
formalism, Class. Quant. Grav. 11 (1994) 2181 and hep-th/9309085.

\bibitem{me2}  D. Anselmi, More on the subtraction algorithm, Class. Quant.
Grav. 12 (1995)\ 319 and hep-th/9407023.

\bibitem{back}  B.S. de Witt, Quantum theory of gravity. II. The manifest
covariant theory, Phys. Rev. 162 (1967) 1195.

\bibitem{call}  D. Kramer, H. Stephani, E. Herlt and M. MacCallum, {\it %
Exact solutions of Einstein's field equations}, Cambridge University Press,
Cambridge, 1980.

\bibitem{dan}  E. D'Hoker, D.Z. Freedman, S.D. Mathur, A. Matusis and L.
Rastelli, Graviton and gauge boson propagators in AdS$_{d+1}$, Nucl. Phys. B
562 (1999) 330 and hep-th/9902042.

\bibitem{turyn}  M. Turyn, The graviton propagator in maximally symmetric
spaces, J. Math. Phys. 31 (1990) 669.
\end{thebibliography}
\end{document}